\documentclass[iop]{emulateapj}
\usepackage{graphicx,color}
\usepackage{natbib}
\usepackage{xcolor}

\begin{document}

\title{XO-2b: a hot Jupiter with a variable host star that potentially affects its measured transit depth}

\author{Robert T. Zellem}
\affil{Department of Planetary Sciences, Lunar and Planetary Laboratory, University of Arizona, 1629 East University Boulevard, University of Arizona, Tucson, AZ, 85721, USA; rzellem@lpl.arizona.edu}

\author{Caitlin A. Griffith}
\affil{Department of Planetary Sciences, Lunar and Planetary Laboratory, University of Arizona, 1629 East University Boulevard, University of Arizona, Tucson, AZ, 85721, USA; griffith@lpl.arizona.edu}

\author{Kyle A. Pearson}
\affil{Department of Astronomy, Steward Observatory, University of Arizona, 933 North Cherry Avenue, Tucson, AZ, 85721,
Currently Department of Physics and Astronomy, Northern Arizona University, Flagstaff, AZ, 86001, USA}

\author{Jake D. Turner}
\affil{Department of Planetary Sciences, University of Arizona, 933 North Cherry Avenue, Tucson, AZ, 85721, USA,
Currently Department of Astronomy, University of Virginia, Charlottesville, VA, 22904, USA} 

\author{Gregory W. Henry}
\affil{{{Center of Excellence in Information Systems,}} Tennessee State University, 3500 John A. Merritt Blvd., PO Box 9501, Nashville, TN 37209, USA} 

\author{Michael H. Williamson}
\affil{{{Center of Excellence in Information Systems,}} Tennessee State University, 3500 John A. Merritt Blvd., PO Box 9501, Nashville, TN 37209, USA} 

\author{M. Ryleigh Fitzpatrick}
\affil{Department of Astronomy, Steward Observatory, University of Arizona, 933 North Cherry Avenue, Tucson, AZ, 85721, USA}

\author{Johanna K. Teske}
\affil{Department of Astronomy, Steward Observatory, University of Arizona, 933 North Cherry Avenue, Tucson, AZ, 85721, USA,
Currently Carnegie DTM, 5241 Broad Branch Road, NW, Washington, DC 20015, USA, Carnegie Origins Fellow, jointly appointed by Carnegie DTM \& Carnegie Observatories}

\author{Lauren I. Biddle}
\affil{Department of Astronomy, Steward Observatory, University of Arizona, 933 North Cherry Avenue, Tucson, AZ, 85721, USA,
Currently Lowell Observatory, 1400 West Mars Hill Road, Flagstaff, AZ 86001, USA}

%\author{Richard S. Freedman}
%\affil{NASA Ames Research Center, Mail Stop 245-3, Moffett Field, CA 94035-1000}

\begin{abstract}
The transiting hot Jupiter XO-2b is an ideal target for multi-object photometry and spectroscopy as it has a relatively bright ($V$-mag = 11.25) K0V host star (XO-2N) and a large planet-to-star contrast ratio (R$_{p}$/R$_{s}\approx0.015$). It also has a nearby (31.21'') binary stellar companion (XO-2S) of nearly the same brightness ($V$-mag = 11.20) and spectral type (G9V), allowing for the characterization and removal of shared systematic errors (e.g., airmass brightness variations). We have therefore conducted a multiyear (2012--2015) study of XO-2b with the University of Arizona's 61'' (1.55~m) Kuiper Telescope and Mont4k CCD in the Bessel U and Harris B photometric passbands to measure its Rayleigh scattering slope to place upper limits on the pressure-dependent radius at, e.g., 10~bar. Such measurements are needed to constrain its derived molecular abundances from primary transit observations. We have also been monitoring XO-2N since the 2013--2014 winter season with Tennessee State University's Celestron-14 (0.36~m) automated imaging telescope to investigate stellar variability, which could affect XO-2b's transit depth. Our observations indicate that XO-2N is variable, potentially due to {{cool star}} spots, {{with a peak-to-peak amplitude of $0.0049 \pm 0.0007$~R-mag and a period of $29.89 \pm 0.16$~days for the 2013--2014 observing season and a peak-to-peak amplitude of $0.0035 \pm 0.0007$~R-mag and $27.34 \pm 0.21$~day period for the 2014--2015 observing season. Because of}} the likely influence of XO-2N's variability on the derivation of XO-2b's transit depth, we cannot bin multiple nights of data to decrease our uncertainties, preventing us from constraining its gas abundances. This study demonstrates that long-term monitoring programs of exoplanet host stars are crucial for understanding host star variability.

%Despite the high correlation between XO-2N's activity and the measured 61'' Kuiper Telescope transit depths, all of the depths agree to $\sim1\sigma$, suggesting that XO-2N's variability influences the light curves below the precision of the 61''. 

%Thus we cannot constrain XO-2b's transit depth beyond 11.72\% in the U-band and 11.17\% in the B-band
\end{abstract}

%We predict that for larger telescopes ($\gtrsim$3~m), XO-2N's variability must be taken into account when measuring XO-2b's transit depth at visible wavelengths. 


%% From the front matter, we move on to the body of the paper.
%% In the first two sections, notice the use of the natbib  \cite
%% and \citet commands to identify citations.  The citations are
%% tied to the reference list via symbolic KEYs. The KEY corresponds
%% to the KEY in the \bibitem in the reference list below. We have
%% chosen the first three characters of the first author's name plus
%% the last two numeral of the year of publication as our KEY for
%% each reference.

\section{Introduction}

The exoplanet XO-2b is arguably typical of the more extensively measured transiting exoplanets, 
considering its basic characteristics and available observations.  It is a Jupiter-sized body that orbits a K0V star at a distance of 0.0369$\pm$0.002 AU and a period of 2.6 days \citep{Burke07}.  The relatively bright \cite[$V$-mag = 11.25;][]{Benavides10} host star, XO-2N, has a southern binary companion, XO-2S, that resides too far away (4600 AU) to affect the planetary orbit \citep{Burke07}.   Both stars have relatively high metallicities and C/O ratios \citep{Teske13a}, as well as the same stellar type and brightness. Thus, XO-2S provides an ideal photometric reference star for transit observations of the exoplanet and host star. 

Ground-based observations of XO-2b at optical wavelengths reveal the presence of both sodium and potassium \citep{Sing11,Sing12}.  Broad-band photometry measures the planetary radius at which its atmosphere becomes optically thick in the Sloan $z$ band \citep{Fernandez09} and in the broad 0.4--0.7 $\mu$m band \citep{Burke07}. {{A large, multi-platform, ground-based study of the XO-2 system finds evidence for variability in both XO-2N and XO-2S and records fourteen R-band photometry primary transits of XO-2b \citep{Damasso15}.}} Secondary eclipse $Spitzer$/IRAC observations suggest that XO-2b has a weak thermal inversion \citep{Machalek09}. The {\it Hubble Space Telescope} (HST) with NICMOS obtained a XO-2b transmission spectrum at near-IR wavelengths of 1.2--1.8 $\mu$m, potentially indicating the presence of water vapor in its atmosphere \citep{Crouzet12}. 

Molecular abundances are difficult to determine from near-IR primary transit 
measurements because the derived mixing ratios depend on the radius assumed as a function of pressure level \citep{Tinetti10,Benneke12,Benneke13}.
\citet{Griffith14a} finds that the extinction coefficient depends 
exponentially on the pressure-dependent radius, indicating that 
high precision measurements of exoplanetary radii are essential for the 
interpretation of transit spectra.
Note that the derived mixing ratios are not sensitive, separately, 
to uncertainties in the host star's radius, because
primary transits measure the ratio of the planet-to-star radius \citep{Griffith14a}. 
Thus the derived planet's radius scales to the assumed stellar radius, 
which we take to be 0.964 $R_{Sun}$ for the 
host star XO-2N.
Gas abundances determined from primary transit data are also sensitive to the 
temperature profile of the planet, but to a much lesser extent  \citep{Griffith14a}. 

Here we present repeated ground-based primary transit 
measurements of XO-2b's atmospheric transmission with the University of Arizona's 61'' (1.55~m) Kuiper Telescope in the Harris B (330--550 nm) and
Bessell U (303--417 nm) photometric bands, where 
the spectrum is devoid of molecular features. 
If the exoplanet is cloudless, the opacities within these bands are established by H$_{2}$ Rayleigh scattering, 
and thus the atmospheric density structure and mean molecular weight, 
rather than atomic and molecular features.  The mean molecular weight is already 
well known through XO-2b's density \citep{Narita11,Burke07}, which indicates a H$_2$-He 
based atmosphere.  Constraints on XO-2b's thermal profile are provided by previous $Spitzer$/IRAC observations \citep{Machalek09}. If clouds are present, an upper 
limit to the radius is obtained, as further discussed below.  

%However, we can only place a weak upper limit on XO-2b's radius due to variability in its host star XO-2N. This interpretation is supported by an extensive monitoring program, 
%the observations of which are also presented here. This variability is likely the result of star spot activity and can influence XO-2b's transit depth \citep[e.g.][]{Pont08, Agol10, Carter11,Desert11,Sing11b}, preventing us from binning data from multiple nights in order to decrease our overall uncertainties.

Our study also includes a nightly monitoring program of the host star XO-2N for variability with Tennessee State University's {{C14 (0.36~m) Automatic Photoelectric Telescope (AIT)}} at Fairborn Observatory \citep{Henry99}. These R-band photometric measurements suggest that XO-2N is indeed variable, likely the result of star spot activity, and could influence XO-2b's transit depth {{\citep[e.g.][]{Pont08, Agol10, Carter11,Desert11,Sing11,McCullough14,Oshagh14,Damasso15}}}. However, it is unclear how much this variability alters XO-2b's signal as all our Kuiper/Mont4k measurements agree to $\sim$1$\sigma$. Regardless, XO-2N's variability prevents us from binning the multiple years of Kuiper/Mont4k data in order to achieve a higher precision measurement of XO-2b's planet-to-star radius ratio. As a result, we cannot strongly constrain its molecular abundances.

%\textbf{Indicate the result here... so as to not mislead the reader
%We find that the XO-2b exhibits variability. 
%This interpretation is supported by an extensive monitoring program, 
%the observations of which are also presented here. }

\section{Observations and Data Reduction} 

Photometric measurements were conducted at the University of Arizona's 61'' (1.55~m) Kuiper Telescope, equipped with  the Mont4k CCD,  a 4096$\times$4096 pixel sensor with a 9.7'$\times$9.7' field of view. Bessel U-band measurements alone were obtained on 2012 January 5, 2013 January 21, and 2013 February 24 (UT). Harris B-band measurements alone were obtained on 2012 October 29 and 2014 January 30 (UT). Simultaneous U and B-band measurements were taken on 2012 December 10, 2014 February 12, 2015 January 18, and 2015 February 8 (UT). Typical seeing was $\sim$1.5''. At the telescope, we binned the pixels by $3 \times 3$ to shorten the readout time and achieve a plate scale of 0.43''/pixel. To ensure accurate time-keeping, an on-board clock was automatically synchronized with GPS every few seconds throughout the observational period.

To help characterize the out-of-transit baseline, each set of observations began $\sim$1--2 hours prior to transit ingress and ended $\sim$1--2 hours after egress. The flux from XO-2N was measured simultaneously with the flux from 8 additional stars on the CCD. Each image is bias-subtracted and flat-fielded with 10 flats and bias frames, using standard IRAF reduction procedures  \citep{Tody93}.  Although 8 reference stars were imaged, only the binary companion, XO-2S, is used as a comparison star to map out systematic errors (e.g., airmass), because it is nearby (31.21'') and of similar stellar type \citep[G9V;][]{Benavides10} and brightness \citep[$V$-mag$=11.20$;][]{Hog00} to the host star. 
In addition, we distinguish a short periodicity in one of the stars within the field of view (RA: 7:48:05.44; Dec: +50:15:56.1), thereby ruling it out as a reference star.

%Light curves are derived from a sequence of observations that span roughly 1--2 hours before the transit to 1--2 hours following the transit, adopting the procedures described in \cite{Pearson14}.

To extract the time-varying flux of the target and comparison stars, we use the Exoplanet Data Reduction Pipeline \citep[ExoDRPL,][]{Pearson14}. Aperture radius sizes ranging from 7 to 14 pixels (0.98'' to 1.96'') with steps of 0.1 pixel (0.014'') are explored to incorporate sufficient flux from each star and ensure that no contaminating light from nearby stars is present in the radial profile. A constant sky annulus with an inner and outer radius of 16 and 20 pixels (2.24'' and 2.8''), respectively, both larger than the target aperture, prevent the inclusion of any background star light. The aperture radii that produce the lowest scatter in the raw XO-2N light curves {{were used}} to create the final light curves. This selection process minimizes both the point-to-point scatter as well as airmass effects. For each night, the light curves of XO-2N are then divided by the light curves of XO-2S to remove shared systematic errors, the largest of which is airmass.

%All the data in the out-of-transit baseline have a photometric RMS between 1-3 mmag, which is typical for the Mont4k on the 1.55 meter Kuiper telescope for high S/N transit photometry \citep{Teske13b,Turner13,Dittmann09a,Dittmann09b,Dittmann10,Dittmann12,Scuderi10}.

%\subsection{Light Curve Extraction} 

The light curves are derived from the data with the analytic equations of \cite{Mandel02} to generate a model transit. 
A Levenberg-Marquardt (LM) non-linear least squares minimization \citep{Press92} via the Interactive Data Language \texttt{mpfit} function \citep{Markwardt09} provides an initial local fit of the model light curves to our data. During the entire analysis, the time of mid-transit (${T_{c}}$), planet-to-star radius ratio R$_{p}$/R$_{s}$, and out-of-transit normalized flux $a$ are left as the only free parameters. The orbital period, inclination, scaled semi-major axis, eccentricity, and quadratic limb darkening coefficients remained fixed for this analysis and are listed in Table~\ref{table:fixedparams}. In cases where the reduced chi-squared ($\chi^{2}_{r}$) of the data to the initial-fit model is greater than unity, we multiply the photometric error bars by $\sqrt{\chi^{2}_{r}}$ to compensate for the underestimated observational errors \citep{Bruntt06,Southworth07} in order to produce a global best-fit solution with a reduced $\chi^{2}$ equal to unity. This inflation technique captures some of the additional non-photon (red) noise {{in}} the data; the time-correlated component of the red noise is captured later in the analysis with a prayer bead.

\begin{deluxetable*}{lcc}
%\tablenum{2}
%\tablecolumns{3}
%\tabletypesize{\footnotesize}
\tablecaption{Fixed Model Values}
%\tablewidth{0pt} 
\tablehead{
\colhead{Parameter}            &
\colhead{Values}            &
\colhead{Reference}           
}
\startdata    
\textbf{XO-2b's Orbital Parameters} & & \\
 Period (days)							&   2.61586178    &  \citet{Sing11}   \\
 Inclination ($\degr$)  					&   88.01             & \citet{Crouzet12}  \\
 a/R$_{s}$   							&   7.986	           & \citet{Crouzet12} \\
 Eccentricity						&   0	& \citet{Crouzet12}		   \\
& & \\
 \textbf{Host Star XO-2N Parameters} & & \\
%T$_{eff}$				&	5340 K			& \citet{Torres08}\\
%$\log{g}$				&	4.452		& \citet{Torres08}\\
%$[Fe/H]$				&	0.450			& \citet{Torres08}\\
T$_{eff}$				&	5343 K			& \citet{Teske13a}\\
$\log{g}$ (cgs)				&	4.49		& \citet{Teske13a}\\
$[Fe/H]$				&	0.39			& \citet{Teske13a}\\
% Omega							&   0			 \\
%Limb darkening coefficients (U-band)	        &  0.98871793, -0.13241752  & \citet{Eastman13} \\
%Limb darkening coefficients (B-band)         &  0.82200365, 0.020254000 & \citet{Eastman13}\\ 
Limb darkening coefficients (U-band)	        &  0.98884840, $-$0.13270563  & \citet{Eastman13} \\
Limb darkening coefficients (B-band)         &  0.81461289, 0.026193820 & \citet{Eastman13}\\ 

\enddata
\label{table:fixedparams}
\tablenotetext{}{{{Please note that the \citet{Teske13a} values used here agree with a more contemporary study by \citet{Damasso15} to 1$\sigma$.}}}
\end{deluxetable*}

To find a global fit solution, we employ 4 simultaneous Markov Chain Monte Carlo \citep[MCMC;][]{Ford05} chains of 200,000 links (steps) seeded by randomly perturbing the best-fit parameters from the LM analysis. Each MCMC chain draws from a prior Gaussian distribution determined by the initial LM fit and tuned so that each fitted parameter has an acceptance ratio of $\sim$0.25 to maximize chain efficiency \citep{Ford05}. After running each MCMC chain, we define the burn-in point for each chain as where the $\chi^{2}$ value first falls below the median of all the $\chi^{2}$ values in the chain, and remove all links prior to this point. A Gelman-Rubin test \citep{GelmanRubin92} finds a scale reduction factor $\le$1.001 for all fitted parameters \citep{Ford05}, suggesting that all chains converge to the same global solutions.

%\textbf{insert stuff about MCMC + Gelman-Rubin here, R$\le$1.0013 for all chains}

\begin{figure}%[htbp]
\begin{center}
\includegraphics[width=0.9\columnwidth]{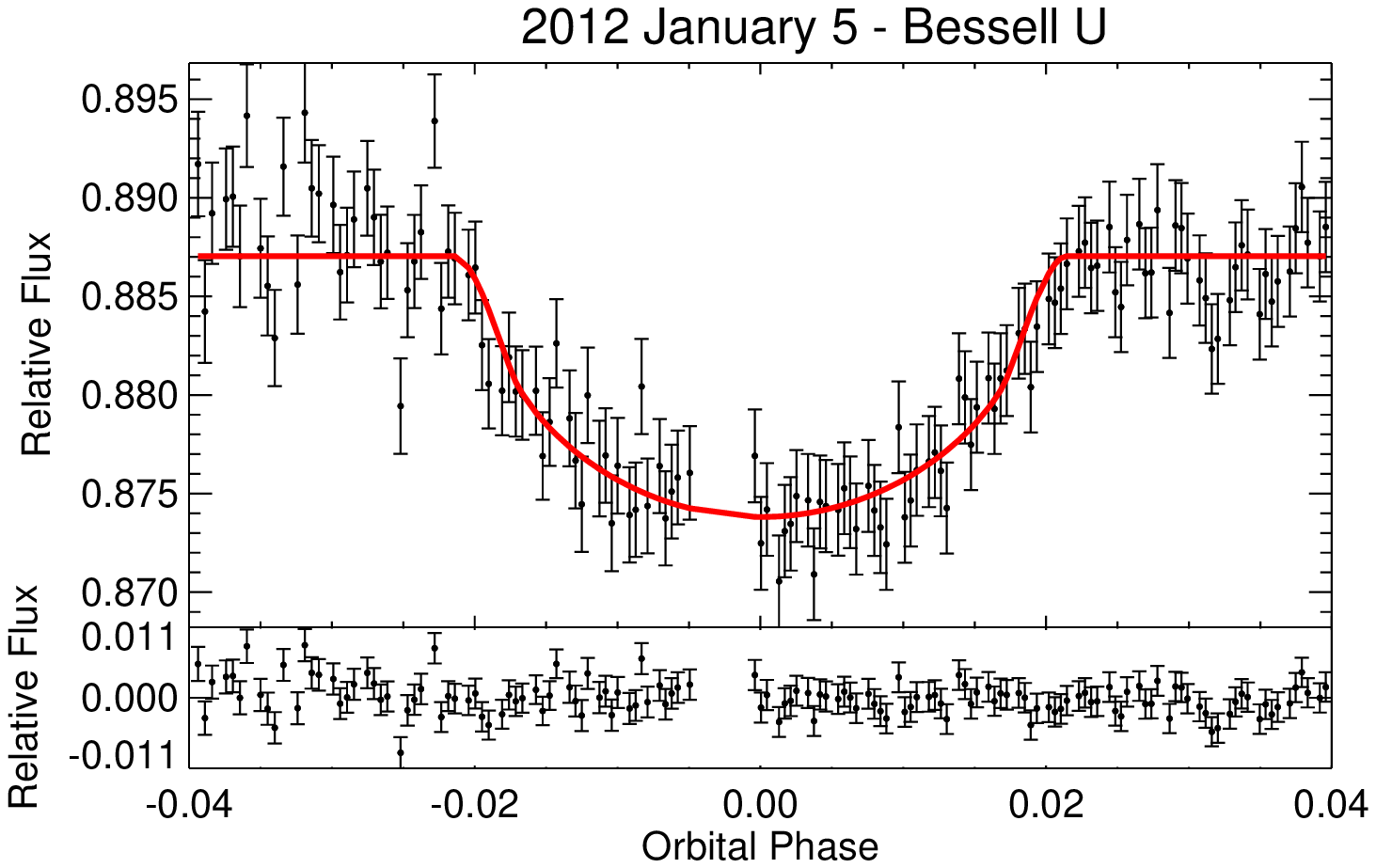}
\includegraphics[width=0.9\columnwidth]{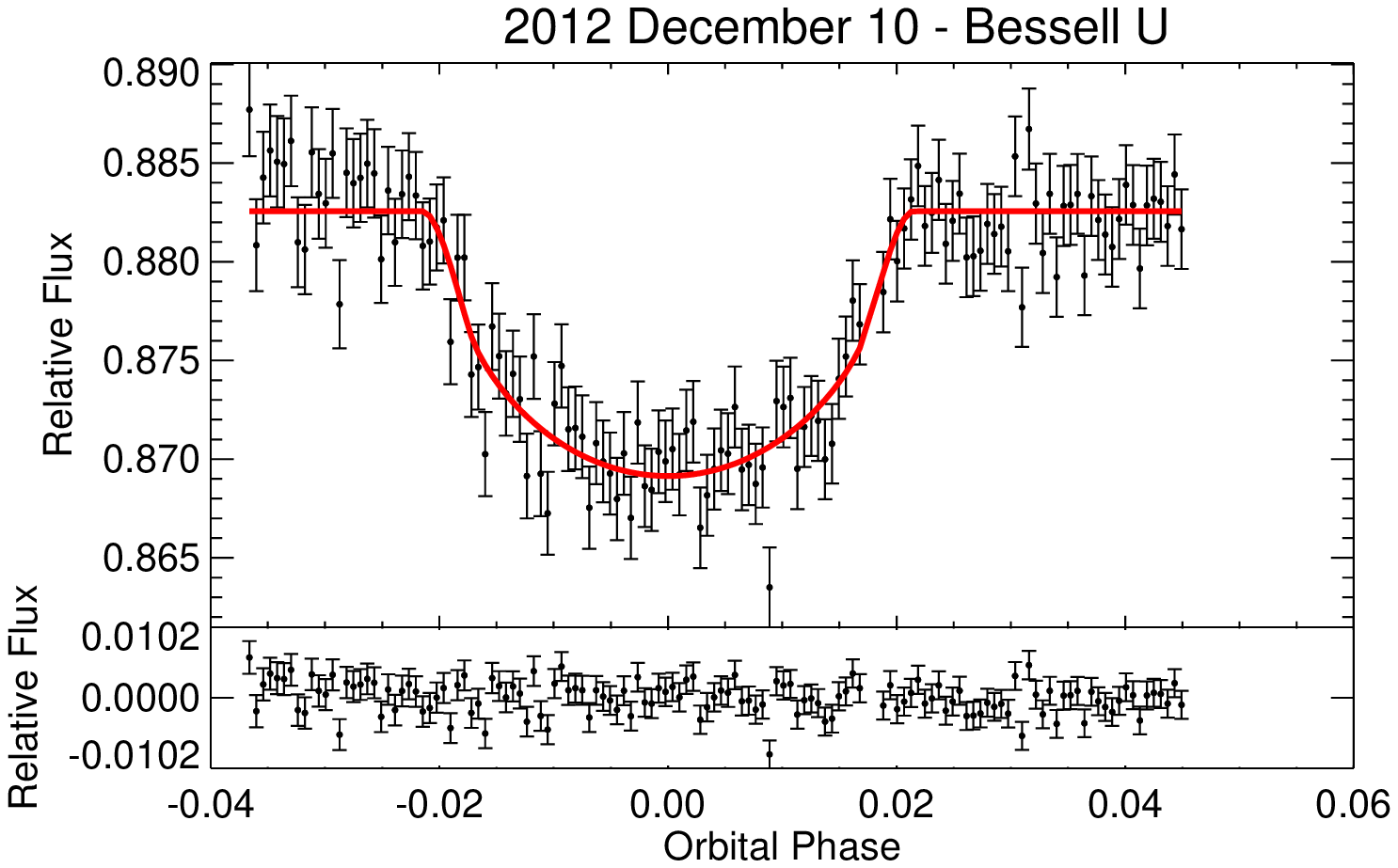}
\includegraphics[width=0.9\columnwidth]{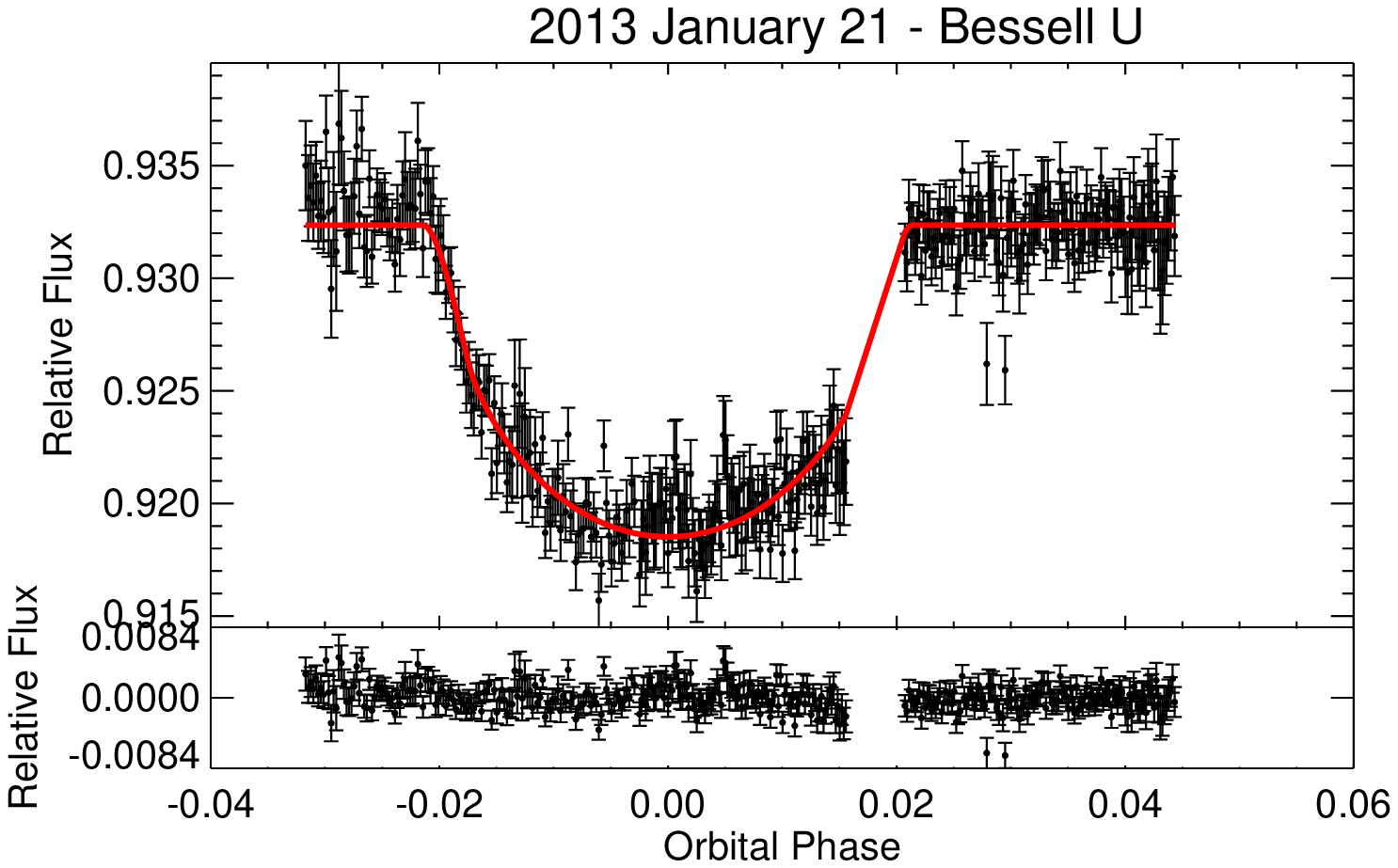}
\includegraphics[width=0.9\columnwidth]{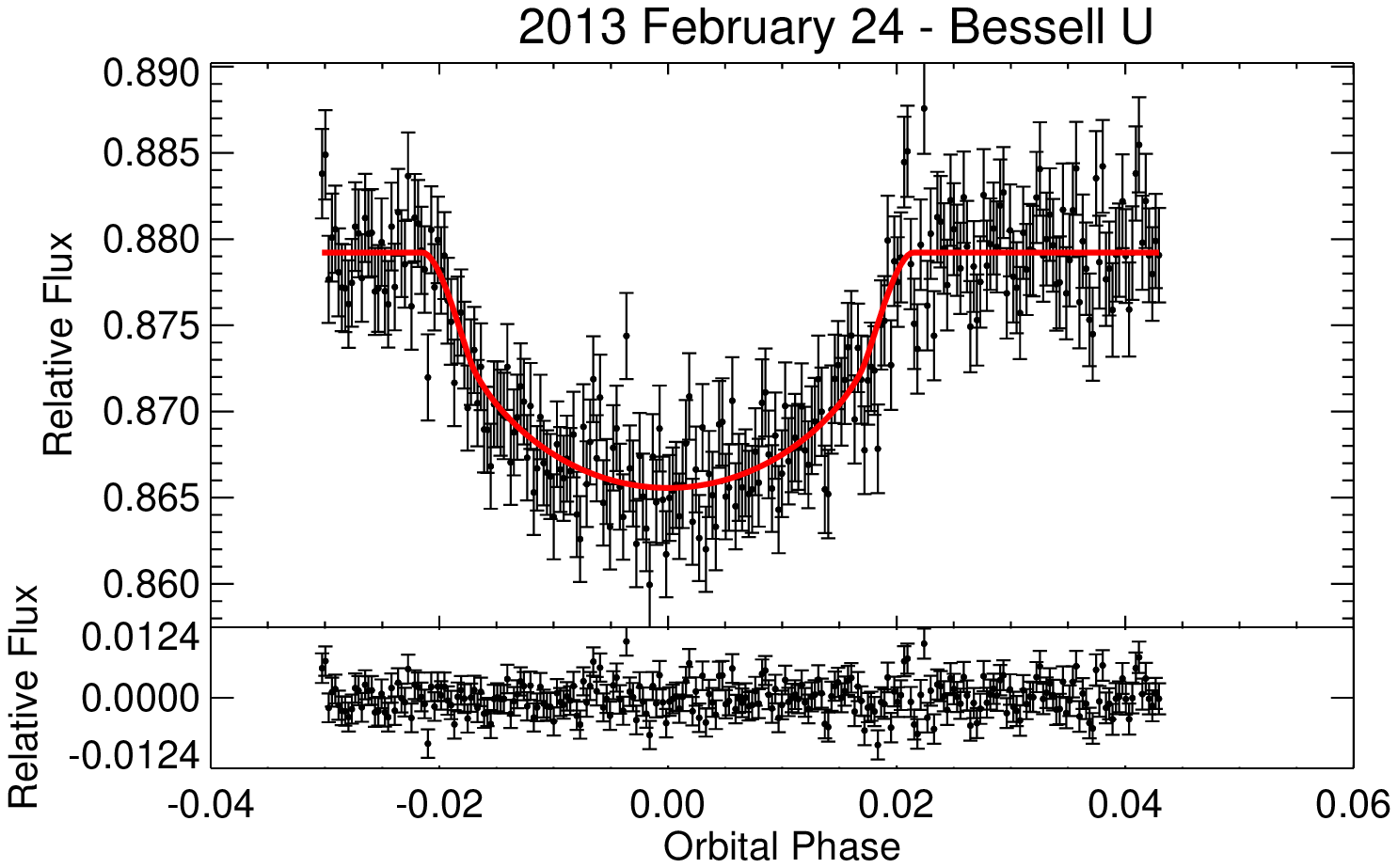}
\caption{Bessell U-band light curves from the 2012--2013 observing seasons showing the transit of XO-2b with their best-fit models indicated with the red line. The 1$\sigma$ error bars include the readout noise, the poisson noise, and the flat-fielding error. The residuals for each light curve are in the lower panel.}
\label{fig:udata1}
\end{center}
\end{figure}

\begin{figure}%[htbp]
\begin{center}
\includegraphics[width=0.9\columnwidth]{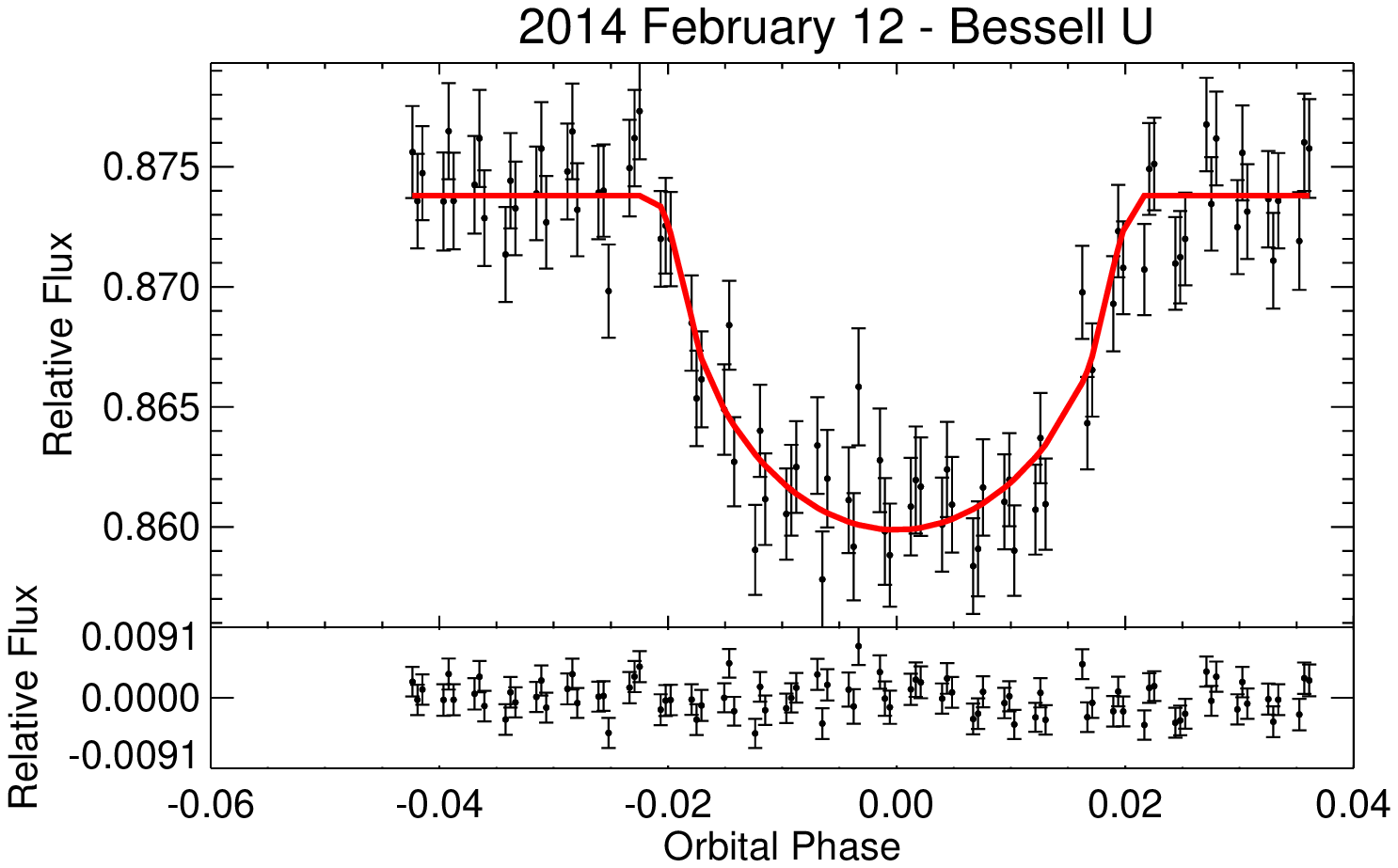}
\includegraphics[width=0.9\columnwidth]{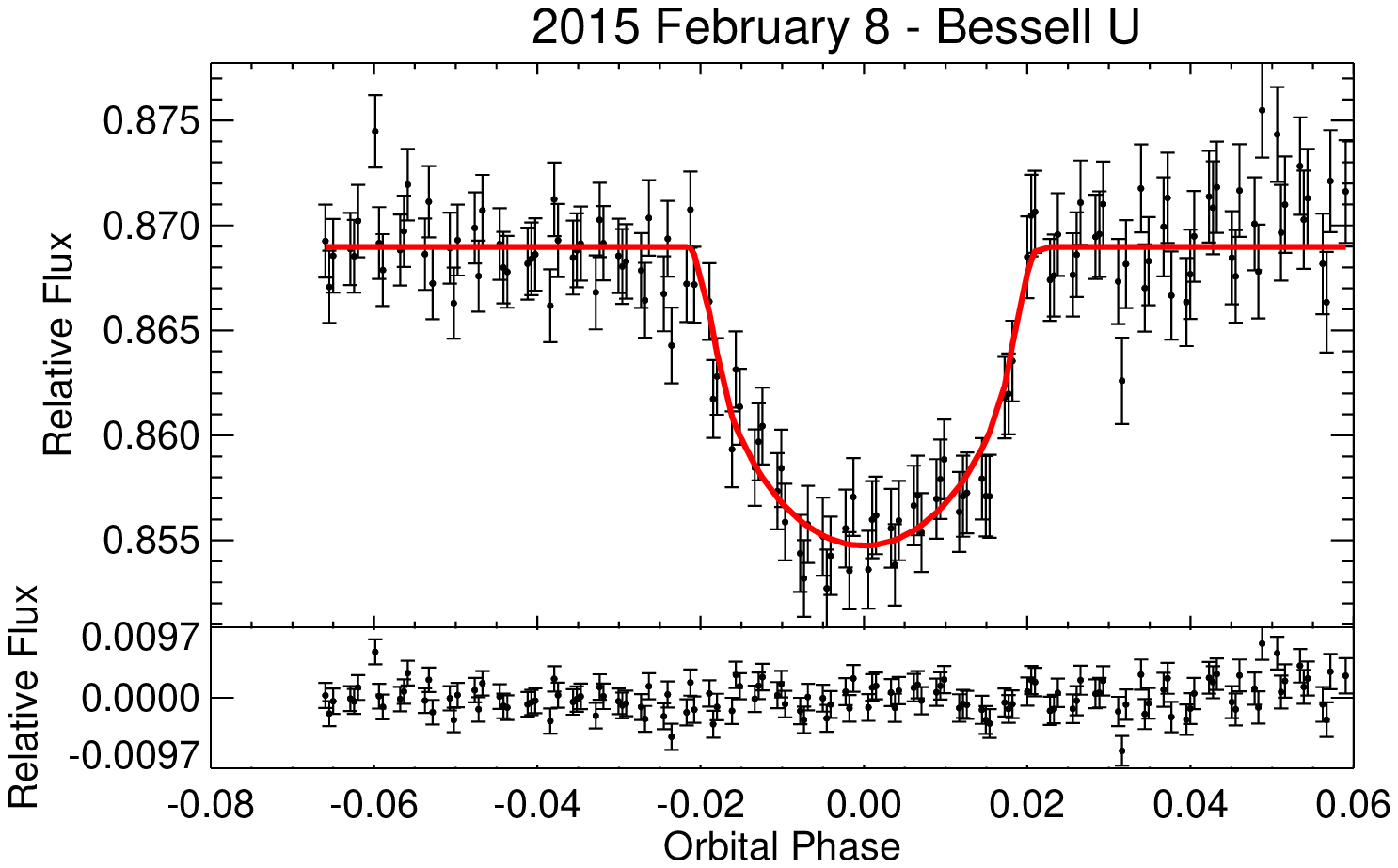}
\includegraphics[width=0.9\columnwidth]{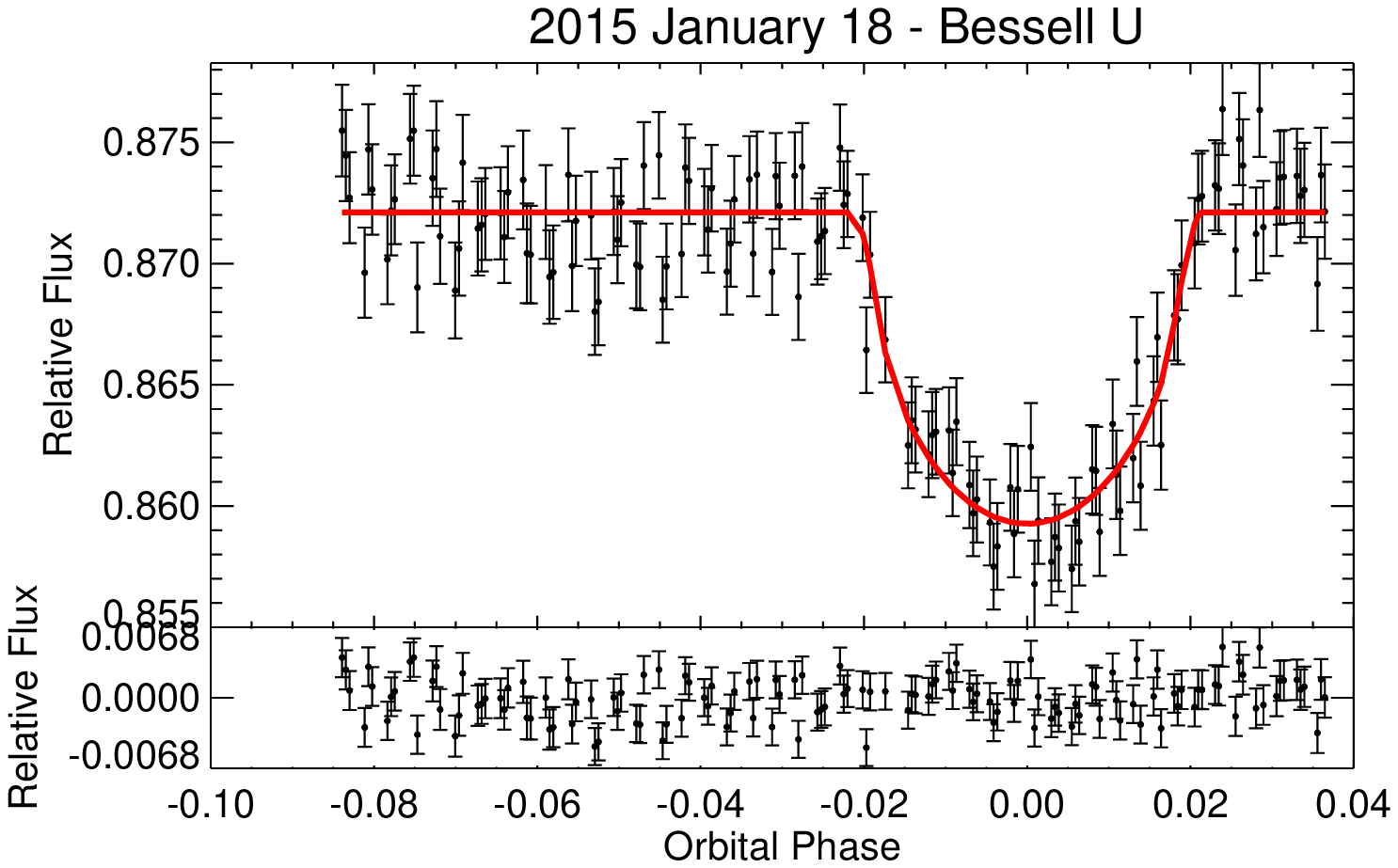}
\caption{Bessell U-band light curves from the 2014--2015 observing seasons showing the transit of XO-2b with their best-fit models indicated with the red line. The 1$\sigma$ error bars include the readout noise, the poisson noise, and the flat-fielding error. The residuals for each light curve are in the lower panel.}
\label{fig:udata2}
\end{center}
\end{figure}

\begin{figure}%[htbp]
\begin{center}
\includegraphics[width=0.9\columnwidth]{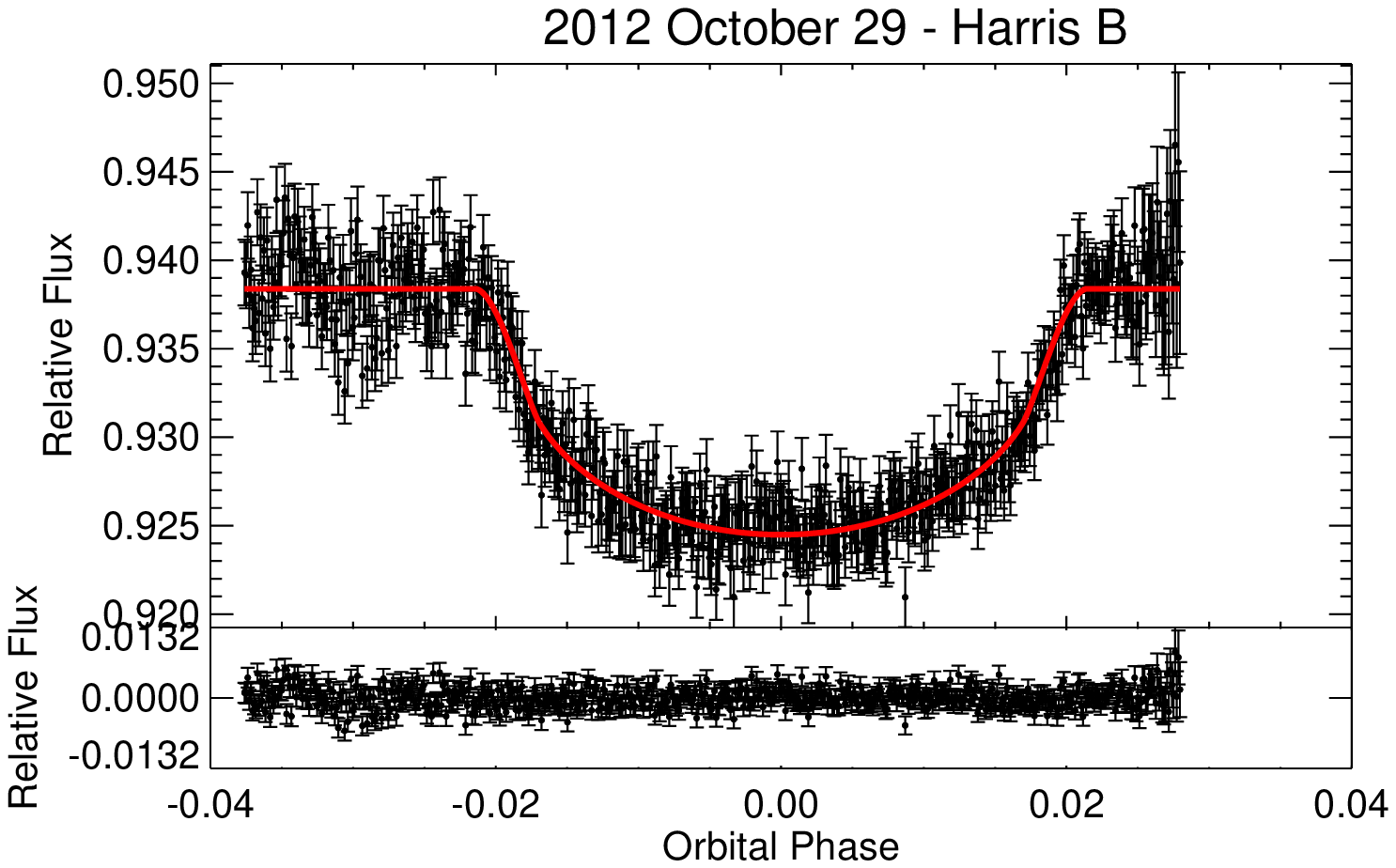}
\includegraphics[width=0.9\columnwidth]{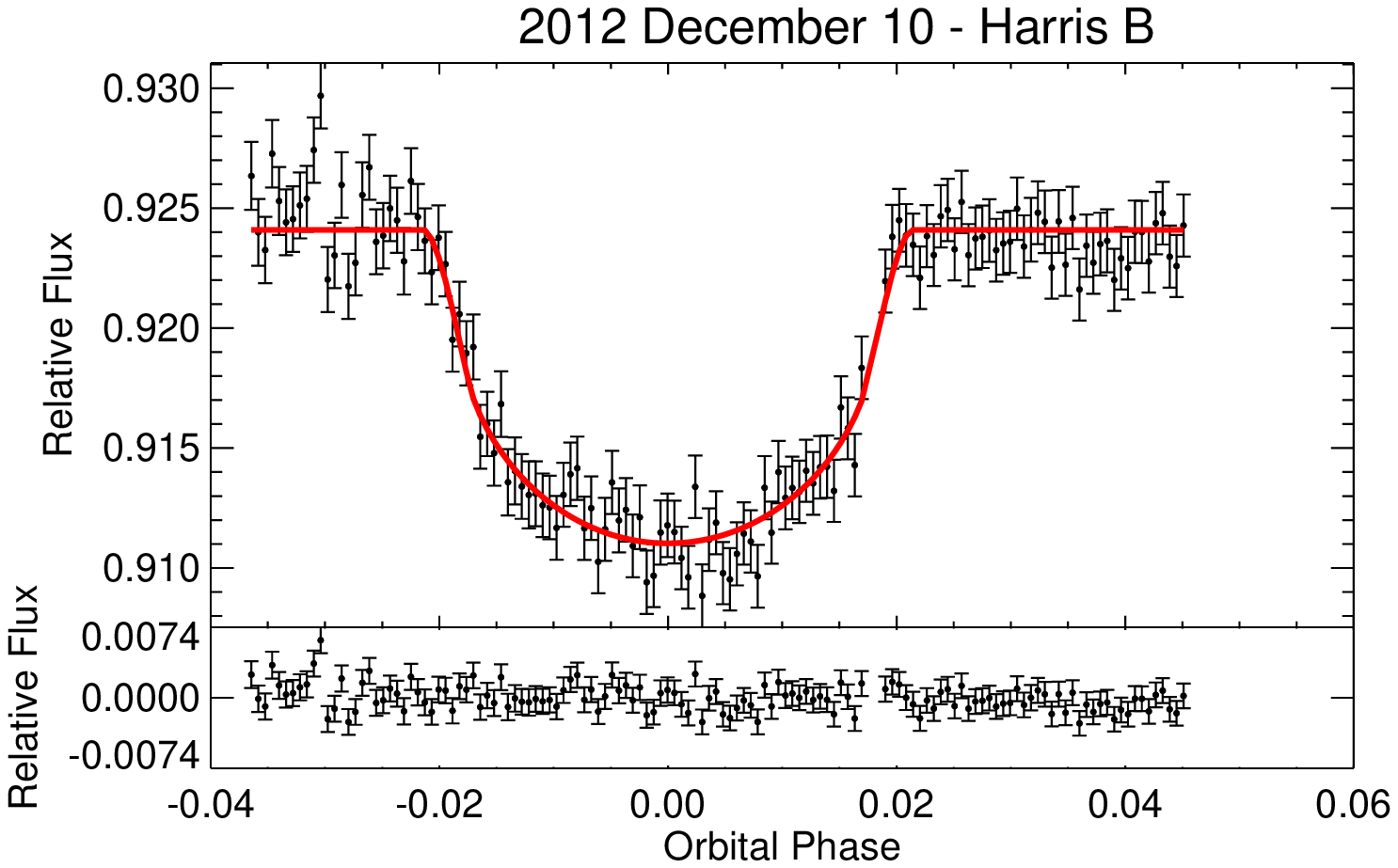}
\includegraphics[width=0.9\columnwidth]{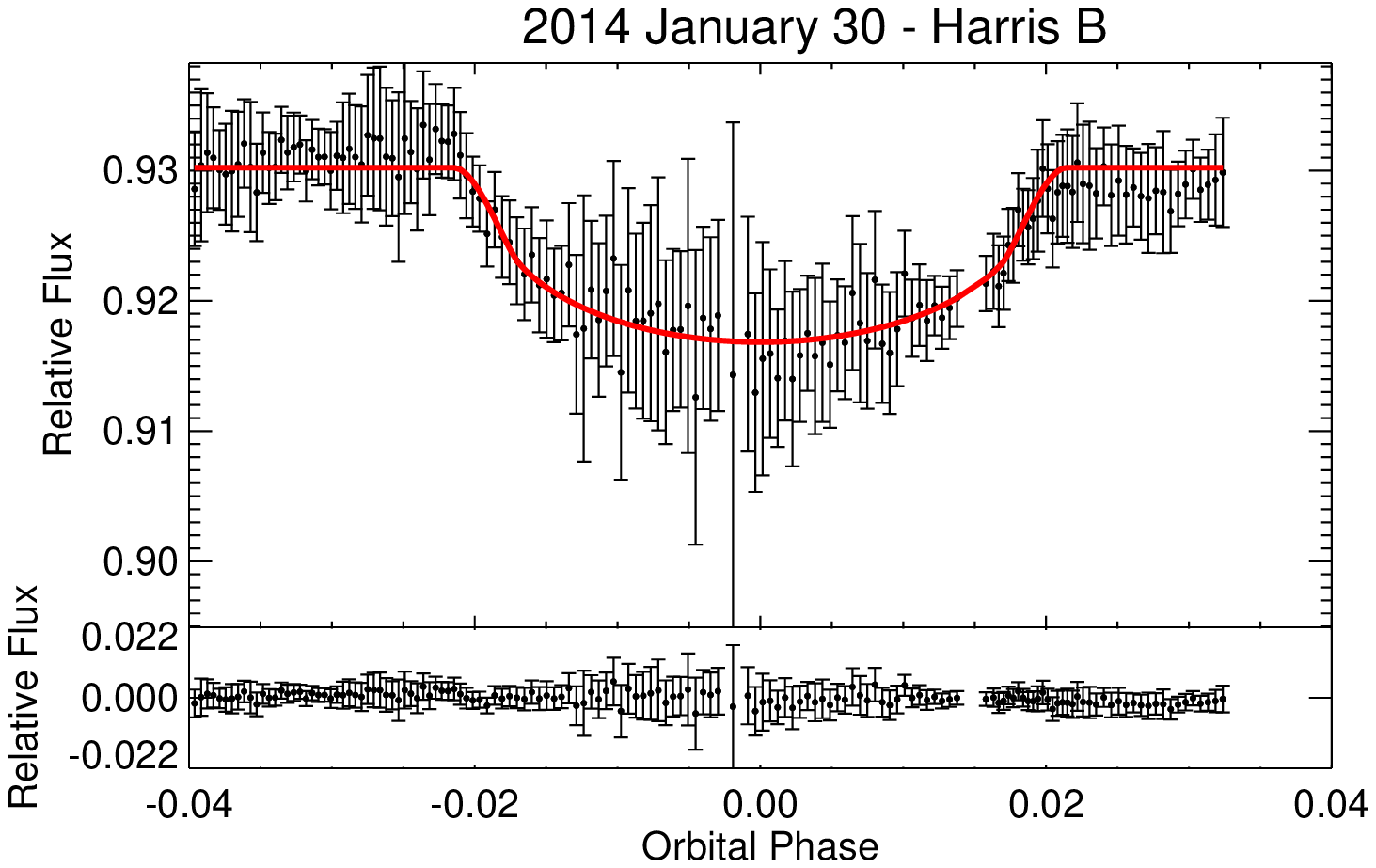}
\caption{Harris B-band light curves from the 2012--2014 seasons showing the transit of XO-2b, following the same protocol as Figure~\ref{fig:udata1}.}
\label{fig:bdata1}
\end{center}
\end{figure}

\begin{figure}%[htbp]
\begin{center}
\includegraphics[width=0.9\columnwidth]{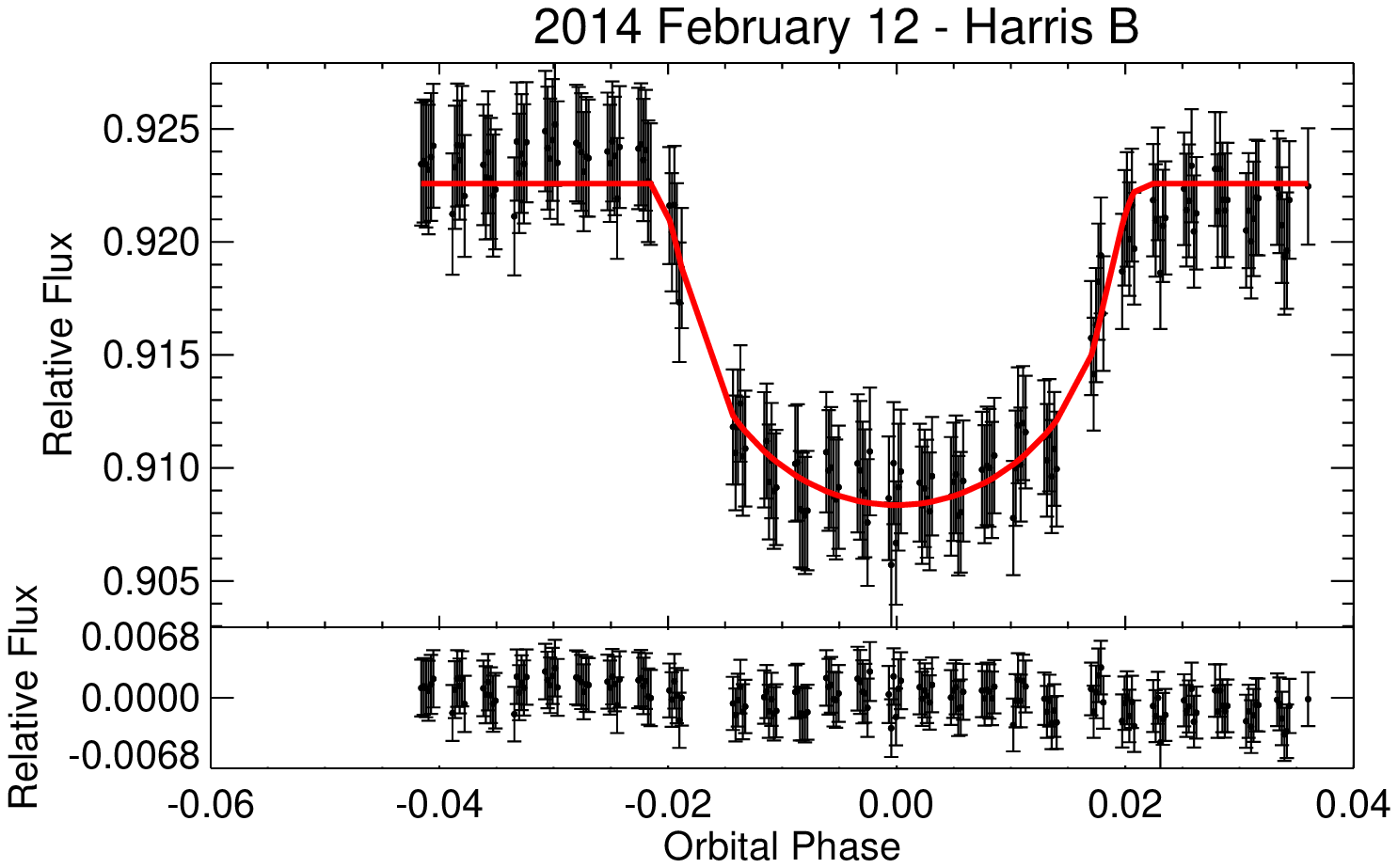}
\includegraphics[width=0.9\columnwidth]{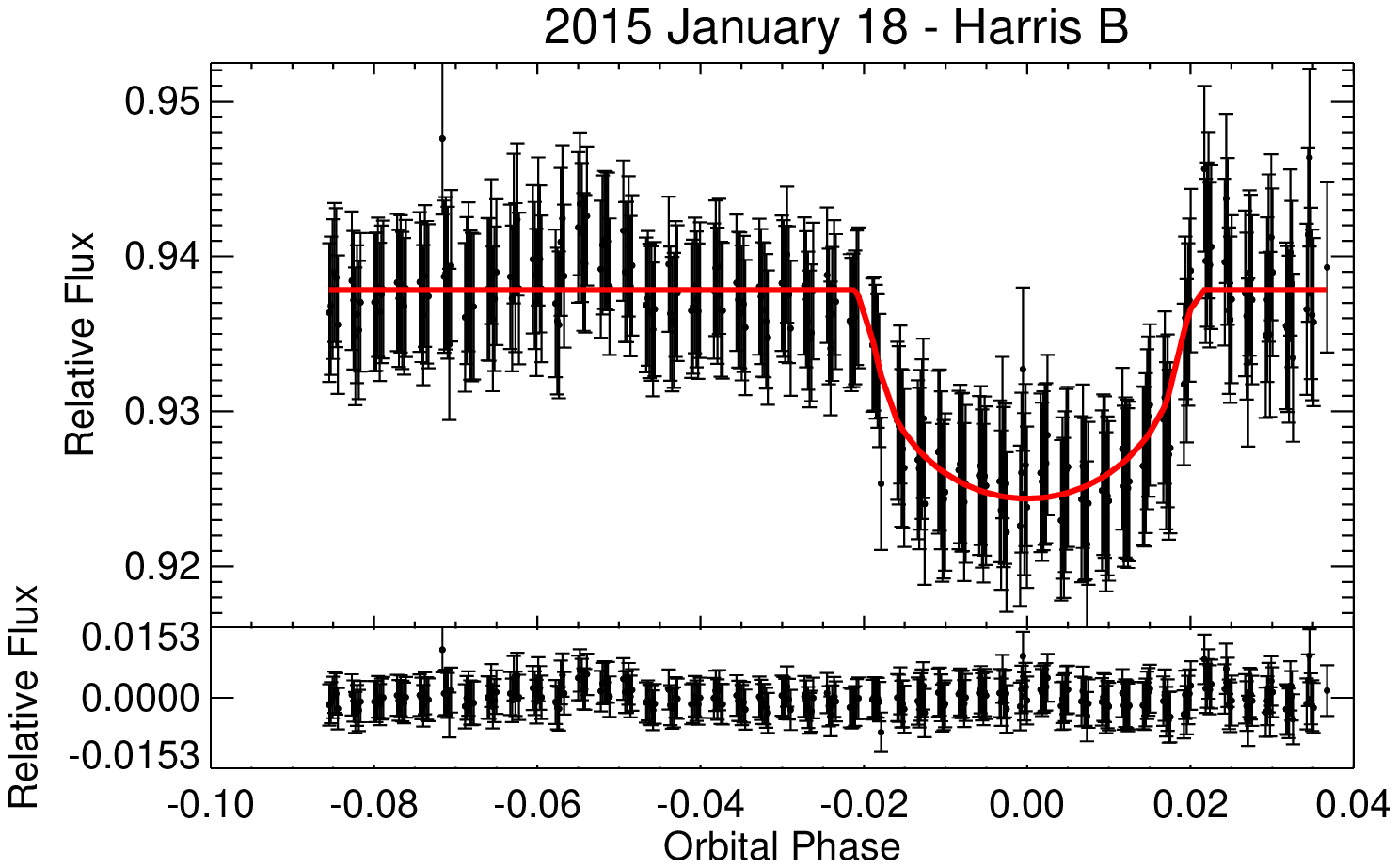}
\includegraphics[width=0.9\columnwidth]{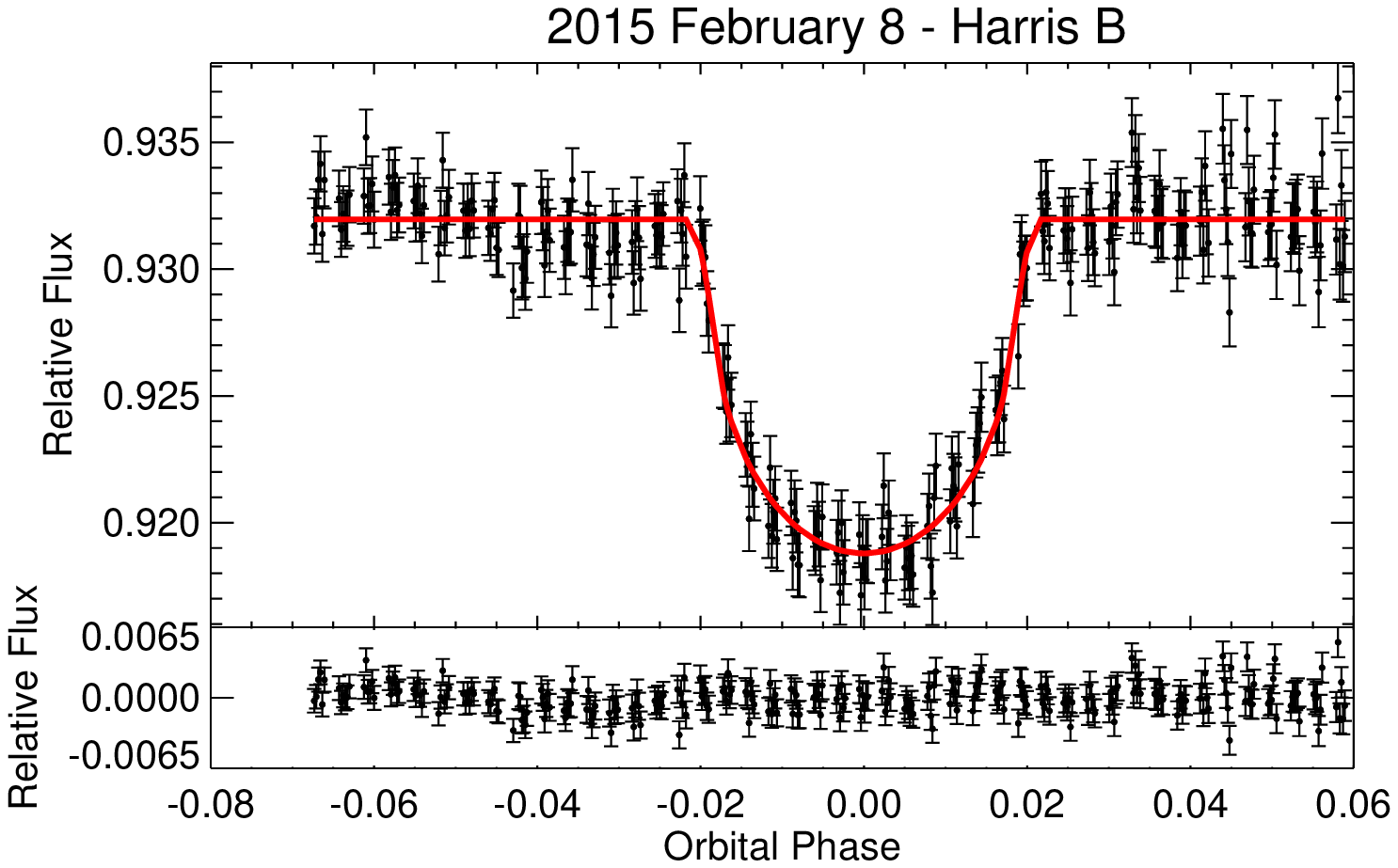}

\caption{Harris B-band light curves from the 2014--2015 seasons showing the transit of XO-2b, following the same protocol as Figure~\ref{fig:udata1}.}
\label{fig:bdata2}
\end{center}
\end{figure}

Since the MCMC draws from a Gaussian prior distribution and assumes that every measurement is uncorrelated, it can underestimate the uncertainties in the derived parameters, particularly if there is non-Gaussian red noise \citep{carterwinn09}. Therefore we use a residual permutation, or ``prayer bead'', method \citep[e.g.,][]{jenkins02, Southworth08, bean08, winn08} to give a more robust estimate {{of}} the parameter uncertainties. The prayer bead method circularly permutes the residuals around the best-fit MCMC solution, and solves for the transit depth, mid-transit time, and out-of-transit normalized flux $a$ with a LM for each permutation. In this way $n$ (where $n$ is the number of data points) new ``simulated'' lightcurves are formed in order to generate new posterior distributions of the parameters. To remain conservative, we choose the larger of the two uncertainties from either the MCMC or the prayer bead as the final error bars for each derived parameter and allow our uncertainties to be non-symmetric \citep[e.g.,][]{Todorov12}. The final model fits to the data are presented in Table~\ref{table:finalparams} and Figures~\ref{fig:udata1}-\ref{fig:bdata2}.

%e also allow
%
%any structure, particularly non-Gaussian red noise, is  The residual permutation involves a series of repeated steps: (1) the best-fit MCMC model is subtracted from the data; (2) the residuals are circularly shifted and added to the data points; (3) a new fit is found with the LM; and (4) the residuals are shifted again, with those at the end wrapped around to the start of the data.  In this way, every new synthetic data set will have the same bumps and wiggles as the actual data, but translated in time. This process is continued until the residuals have cycled back to where they originated. Our resulting parameter values have non-Gaussian distributions, suggesting the presence of red noise. Consequently, we allow our error bars to be non-symmetric, as demonstrated by \cite{Todorov12}.  
%%The final 1-$\sigma$ error bars for the fitted parameters are given in Table~\ref{table:finalparams}.

\begin{deluxetable*}{lcccccccccc}
%\tabletypesize{\scriptsize}
%\tablenum{3}
%\tablecolumns{11}
%\tabletypesize{\footnotesize}
\tablecaption{Derived Parameters}
%\tablewidth{0pt} 
\tablehead{\colhead{\textbf{Date}} & \colhead{\textbf{R$_{p}$/R$_{s}$}} & \multicolumn{2}{c}{\textbf{Uncertainty}} & \colhead{\textbf{Mid-transit time}} & \multicolumn{2}{c}{\textbf{Uncertainty}} & \colhead{\emph{\textbf{a}}\tablenotemark{$\dagger$}} & \multicolumn{2}{c}{\textbf{Uncertainty}} & \colhead{\textbf{$\sigma$/$\sigma_{photon}$}} \\
\colhead{(UT)} & & & & \colhead{($T_{c} - 2456000$ JD)} & & & & & &}
\startdata
%date		Rp/Rs		neg err		pos err		Ephem-2.456d6	neg err		pos err		a		neg err		pos err		photon noise
\multicolumn{11}{l}{\textbf{Bessel U-band}}\\
%2012 Jan 5   &   0.1038 &   $-$0.0034  &  $+$0.0042    &  $-$68.2337  &  $-$0.0029  &  $+$0.0015    &  0.88704  & $-$0.00032 &  $+$0.00041    &   3.62\\
%2012 Dec 10    &  0.1048 &   $-$0.0022 &   $+$0.0029  &     271.8283  &  $-$0.0024 &   $+$0.0015  &    0.88255 &  $-$0.00023 &  $+$0.00025   &    3.21\\
%2013 Jan 21    & 0.1035  &  $-$0.0019 &  $+$0.0023   &    313.6810   & $-$0.0019  &  $+$0.0024   &   0.93236  & $-$0.00014 &  $+$0.00031    &   3.82\\
%2013 Feb 24    & 0.1059  & $-$0.0014 &  $+$0.0015    &   347.6865 &  $-$0.0006  & $+$0.0006  &    0.87922  & $-$0.00024  & $+$0.00023   &    3.88\\
%2014 Feb 12    &  0.1073  &  $-$0.0015  &  $+$0.0024   &    700.8275  &  $-$0.0011  &  $+$0.0018  &    0.87380 &  $-$0.00029 &  $+$0.00027  &     3.36\\
%2015 Jan 18   &  0.1031  & $-$0.0034  &  $+$0.0045    &   1040.8883 &   $-$0.0023  &  $+$0.0039   &   0.87211 &  $-$0.00019  & $+$0.00033   &    2.77\\
%2015 Feb 8    &  0.1088  &  $-$0.0051 &   $+$0.0032   &    1061.8169  &  $-$0.0040  &  $+$0.0023   &   0.86897  & $-$0.00021 &  $+$0.00033   &    3.04\\

2012 Jan 5   & 0.1037  &  $-$0.0035   & $+$0.0041   &   $-$68.2337  &  $-$0.0030  &  $+$0.0016  &    0.88703 &  $-$0.00032 &  $+$0.00041    &   3.71\\
2012 Dec 10    &  0.1047  &  $-$0.0024  &  $+$0.0028   &    271.8283  &  $-$0.0025  &  $+$0.0014  &    0.88254  & $-$0.00024 &  $+$0.00024    &   3.17\\
2013 Jan 21    &  0.1035   & $-$0.0019  &  $+$0.0023    &   313.6810 &   $-$0.0018  &  $+$0.0024   &   0.93235  & $-$0.00014 &  $+$0.00031     &  3.84\\
2013 Feb 24    &  0.1059   & $-$0.0015  &  $+$0.0015    &   347.6865 &  $-$0.0006 &  $+$0.0006  &    0.87921 &  $-$0.00023 &  $+$0.00023   &    3.90\\
2014 Feb 12    &  0.1072  &  $-$0.0020  &  $+$0.0020     &  700.8276  &  $-$0.0012  &  $+$0.0017    &  0.87379  & $-$0.00029  & $+$0.00028    &   3.36\\
2015 Jan 18   &  0.1031  &  $-$0.0035  &  $+$0.0045     &  1040.8883  &  $-$0.0022  &  $+$0.0039   &   0.87210  & $-$0.00019  & $+$0.00033    &   2.79\\
2015 Feb 8    &  0.1087  &  $-$0.0051  &  $+$0.0033     &  1061.8169  &  $-$0.0040  &  $+$0.0023   &   0.86897  & $-$0.00021  & $+$0.00034    &   2.98\\

 & & & & & & & & & & \\
\multicolumn{11}{l}{\textbf{Harris B-band}}\\
%2012 Oct 29 & 0.1054  &   $-$0.0014  &   $+$0.0016   &     229.9736  &  $-$0.0005  &   $+$0.0009   &    0.93839 &   $-$0.00014  &  $+$0.00027   &     3.65\\
%2012 Dec 10 & 0.1030  &   $-$0.0020  &   $+$0.0018   &     271.8288  &   $-$0.0023   &  $+$0.0008    &   0.92410  &  $-$0.00019  &  $+$0.00020   &     3.76\\
%2014 Jan 30 & 0.1039   &  $-$0.0033  &   $+$0.0078    &    687.7504  &   $-$0.0017   &  $+$0.0031    &   0.93023  &  $-$0.00051  &  $+$0.00058   &     11.60\\
%2014 Feb 12 & 0.1075   &  $-$0.0038  &   $+$0.0039   &     700.8291  &   $-$0.0030  &   $+$0.0028    &   0.92258 &   $-$0.00040  &  $+$0.00044  &      8.71\\
%2015 Jan 18 & 0.1037   &  $-$0.0041  &   $+$0.0021   &     1040.8889  &   $-$0.0011  &   $+$0.0035   &    0.93782  &  $-$0.00034  &  $+$0.0032    &    13.82\\
%2015 Feb 8 & 0.1030  &   $-$0.0012  &   $+$0.0028     &   1061.8164  &  $-$0.0007  &   $+$0.0025    &   0.93196  &  $-$0.00010  &  $+$0.00019    &    3.80\\

2012 Oct 29 &  0.1054  &  $-$0.0015  &  $+$0.0016     &  229.9736  & $-$0.0005 &  $+$0.0010   &   0.93839 &  $-$0.00014 &  $+$0.00027    &   3.64\\
2012 Dec 10 &  0.1030  &  $-$0.0021  &  $+$0.0018   &    271.8288  &  $-$0.0023  & $+$0.0008  &    0.92409 &  $-$0.00018  & $+$0.00019   &    3.83\\
2014 Jan 30 &  0.1042   & $-$0.0033  &  $+$0.0075     &  687.7504  &  $-$0.0017  &  $+$0.0031    &  0.93024  & $-$0.00050  & $+$0.00057     &  6.08\\
2014 Feb 12 &  0.1076  &  $-$0.0038  &  $+$0.0039    &   700.8291  &  $-$0.0030  &  $+$0.0028    &  0.92258  & $-$0.00039  & $+$0.00044    &   5.20\\
2015 Jan 18 &  0.1038  &  $-$0.0041  &  $+$0.0023     &  1040.8889  &  $-$0.0012  &  $+$0.0034    &  0.93782  & $-$0.00022  & $+$0.00019     &  6.19\\
2015 Feb 8 &  0.1030  &  $-$0.0012  &  $+$0.0028   &    1061.8164  & $-$0.0007  &  $+$0.0025  &    0.93196  & $-$0.00010 &  $+$0.00019    &   3.71\\

\enddata
\tablenotetext{$\dagger$}{Out-of-transit normalized flux, effectively measures the brightness ratio of the host star XO-2N to the comparison (and binary companion) star XO-2S}
\label{table:finalparams}
\end{deluxetable*}

%%%% -------  TABLE 2

%
%\begin{figure}[htbp]
%\begin{center}
%%\includegraphics[angle=180,scale=0.3]{Best-Fit_05jan12.eps}
%%\includegraphics[angle=90,scale=0.3]{Best-Fit_09dec12_U.eps}
%%\includegraphics[angle=90,scale=0.3]{Best-Fit_23feb13.eps}
%\includegraphics[width=0.5\columnwidth]{final_09Dec2012_U.eps}
%\includegraphics[width=0.5\columnwidth]{final_20Jan2013_U.eps}
%\includegraphics[width=0.5\columnwidth]{final_05Jan2012_U.eps}
%\includegraphics[width=0.5\columnwidth]{final_11Feb2014_U.eps}
%\includegraphics[width=0.5\columnwidth]{final_23Feb2013_U.eps}
%\includegraphics[width=0.5\columnwidth]{final_07Feb2015_U.eps}
%\includegraphics[width=0.5\columnwidth]{final_17Jan2015_U.eps}
%\caption{U band light curves showing the transit of XO-2b. Values are  normalized to one, for each date of observed transit (UTC). Models of the data are shown in red, with residuals shown in the lower panels of each plot. The 1$\sigma$ error bars include the readout noise, the poisson noise, and the flat-fielding error.}
%\label{fig:data1}
%\end{center}
%\end{figure}

%The near-UV light curves do not demonstrate any 
%asymmetries, which are predicted to be present as a result of extra absorption from a bow shock in front of the planet created by interactions between the stellar coronal material and the planet's magnetosphere \citep{Vidotto11}.  This finding is consistent with \cite{Turner13} and \cite{Pearson14} who also detect no asymmetry in the light curve of the transiting exoplanets TrES-3b and HAT-P-16b, respectively. 

%\textbf{Need to make RpRs activity plots into RpRs2; need to fit these with MCMCs rather than just LMs, cannot estimate blackbody curves of the starspots as we do not know the size of the spots}

%\textbf{lack of varying RpRs measurements seemed to reinforce previous studies that XO-2N is a non-active system (Burke, Sing, Crouzet). however, initial analyses }

%\textbf{We find XO-2N is fainter than XO-2S by 0:07 $\pm$ 0:008 mag in the B band, 0:055 $\pm$ 0:004 mag in the V band, 0:040 $\pm$ 0:004 mag in the R band, and 0:030 $\pm$ 0:003 mag in the I band.}

\section{Effects of Stellar Variability}

Initial analyses of the 2012 U and B-band data indicated 1-$\sigma$ agreement between the derived $R_{p}/R_{s}$ values. This consistency seemed to reinforce previous studies that concluded that XO-2 is a non-active system. For example, \citet{Burke07} found that all of their Extended Team XO-2b light curves agreed to $\sim$0.5\%. In addition, \citet{Sing11} found no evidence for occulted starspots in their GTC/OSIRIS spectrophotometric study of XO-2b. However we found that our out-of-transit baseline (as indicated by the out-of-transit normalized flux $a$; see Table~\ref{table:finalparams}) varied 0.6\% in the U and 2.4\% in the B, suggesting that the host star XO-2N is variable. For comparison, \citet{Burke07} found that XO-2N was 93.76\% (0.07 $\pm$ 0.008 mag) the brightness of XO-2S in the B-band, whereas we find that XO-2N varies between 92.258\% and 93.839\% of the brightness of XO-2S.

%While we do not see any conclusive evidence for starspot crossings in any of our light curves, unocculted starspots can cause a transit depth to deepen, resulting in a retrieval of an incorrect R$_{p}$/R$_{s}$ measurement and an incorrect interpretation of the exoplanet's atmosphere \citep[e.g.,][]{Pont08, Agol10, Carter11,Desert11,Sing11b}. In order to investigate this
%possibility, we have monitored XO-2N since the 2013--2014 winter season with Tennessee
%State University's 0.36 and 0.61~meter Automatic Photoelectric Telescopes (AITs) at Fairborn Observatory \citep{Henry99}. In the 2013--2014 season alone, we obtained 140 
%good nights of R-band observations (Figure~\ref{fig:monitor}). A frequency spectrum of these measurements indicate 
%that XO-2N's brightness varies periodically with a $29.9 \pm 0.2$ day period. A least squares sine fit to the data at that period shows a peak-to-peak amplitude of $0.00455 \pm 0.00090$ mag, i.e., a 5.0$\sigma$ detection, indicating that XO-2N's variability is real.  The period and implied stellar rotation is consistent with the 
%29--44 days indicated by the vsin(i) value.  \textbf{(Greg- Which value is this? And what is the citation?)} The variation of XO-2N's brightness is consistent with the presence of starspots.

While we do not see any conclusive evidence for starspot crossings in any of our light curves, unocculted starspots can cause a transit depth to deepen, resulting in the retrieval of an incorrect R$_{p}$/R$_{s}$ measurement and an incorrect interpretation of the exoplanet's atmosphere \citep[e.g.,][]{Pont08, Agol10, Carter11,Desert11,Sing11b}. {{To investigate the possibility of photometric variability in XO-2N and XO-2S, we made nightly observations of both components throughout the 2013--2014 and 
2014--2015 observing seasons with the Tennessee State University Celestron-14 
(C14) Automated Imaging Telescope (AIT) at Fairborn Observatory in southern 
Arizona \citep[e.g.,][]{Henry99,ehf2003}. We acquired $\sim$170 and $\sim$100 
nightly photometric measurements in the Cousins R-band during the two observing seasons.  We 
compute differential magnitudes of XO-2N and XO-2S with respect to the mean 
brightness of several of the most stable comparison stars in the CCD field of 
view and then perform separate frequency analyses of XO-2N and XO-2S
for each observing season (Table~3 and Figs.~5--8). 
A few outliers were removed in each analysis, based on 3-$\sigma$ filtering. 
Further details of our C14 data acquisition and reduction procedures can be 
found in the paper by \citet{sing+2015}, which describes a similar analysis 
of the exoplanet-host star WASP-31.}}

%In order to investigate this possibility, we monitored XO-2N throughout the 2013--2014 observing season with Tennessee State University's Celestron-14 (0.36~m) Automated Imaging Telescope (AIT) at Fairborn Observatory in southern Arizona \citep{Henry99}. Between 2013 October 7 and 2014 May 31, we acquired 171 good R-band observations of XO-2N (Figure~\ref{fig:monitor}). A frequency spectrum reveals low-amplitude brightness variability in XO-2N with a period of $29.9 \pm 0.16$ days.  A least-squares sine fit to the data at that period shows a peak-to-peak amplitude of $0.00492 \pm 0.00070$~mag or 0.53\%, a 7.0$\sigma$ detection. We take this to be the stellar rotation period made visible by the rotational modulation of starspots across the face of the star.  This is consistent with the period range of 29--44 days predicted from the stellar radius of 0.964R$_{\sun}$ adopted in this paper and the $v\sin{i} = 1.4 \pm 0.3$ from the discovery paper of \citet{Burke07}.

{{The 2013--2014 observations of the northern component, XO-2N, have more 
scatter than the 2014--2015 season (Table~3 and Figs.~5 \& 6). The 
frequency spectrum in the middle panel of Figure~5 reveals low-amplitude 
brightness variability in XO-2N with a period of $29.9\pm0.16$ days. A 
least-squares sine fit to the phase curve in the bottom panel of Fig.~5 
shows a peak-to-peak amplitude of $0.0049\pm0.0007$ mag, or 0.53\%, a 
7$\sigma$ detection.  We take this to be the rotation period of the 
northern component made visible by the rotational modulation of starspots 
across the face of the star. This period is consistent with the range of 
29--44 days predicted from the stellar radius of $0.97~R_{\sun}$ and 
$v \sin i = 1.4\pm0.3$ from the discovery paper of \citet{Burke07}.
Brightness variation in the northern component is reduced in the 2014--2015
observing season, but the data essentially confirm the results from the
first season.  The 2.5~day difference in the two periods likely results
from spot evolution on the timescale of the rotation period.}}

{{The observations of the southern component, XO-2S, have very low scatter, 
0.0025 and 0.0023 mag in the first and second seasons, respectively (Table~3 and Figs.~7 \& 8).  These
values are consistent with our measurement  precision for a single 
observation, as determined by inter-comparison of the most constant stars 
in the CCD frame.  Period analyses of XO-2S 
reveal no significant periodicity in either observing season.  Our observed  
levels of variability of XO-2N and XO-2S are consistent with their activity 
levels, $\log R^{\prime}_{HK}$, of $-4.91\pm0.01$ and $-5.02\pm0.01$,
respectively, given by \citet{Damasso15}.  However our period analyses
are inconsistent with those of \citet{Damasso15}, despite their use of a similarly-sized telescope (40~cm) relative to our study (36~cm) and our two monitoring programs partially overlapping in the 2013--2014 season. This discrepancy probably arises due to the difference in sampling: \citet{Damasso15} monitored the XO-2 system for 42~nights in the $I$-band over the 2013--2014 season while our program measured it at a much higher cadence over 171~nights in the 2013--2014 season and 97~nights in the 2014--2015 season.}}

Assuming XO-2N's variability is constant over the last {{4}} years of observations\footnote{Please note that since the coverage of star spots varies with time, {{the amplitudes of the variability observed in the 2013--2014 and 2014--2014 seasons are}} not necessarily applicable to all of our transit observations which were recorded in the 2011--2015 winter seasons {{(Table~2)}}.}, we can plot XO-2b's transit depth in each filter vs. XO-2N's $R$-band variability {{across the 2013--2014 and 2014--2015 seasons (Figs.~\ref{fig:variable} and \ref{fig:variable14}). In the U-band, the transit depth has a $-0.60$ correlation with the host star's 2013--2014 variability and a $0.45$ correlation with the host star's 2014--2015 variability. In the B-band, the correlation is $0.19$ with the host star's 2013--2014 variability and $-0.68$ with the host star's 2014--2015 variability. These plots suggest a correlation between the measured transit depth and host star variability.}} However since all of the Kuiper/Mont4k transit depths agree to $\sim$1$\sigma$, we cannot quantify how stellar variability is affecting the derived transit depth with any statistical significance. Since the variation in the measured transit depth is potentially being influenced by the host star's activity, the multiple transit depth measurements cannot be binned together to achieve higher precision. Thus we cannot constrain XO-2b's transit depth beyond 11.72\% in the U-band and 11.17\% in the B-band.

\begin{deluxetable*}{ccccccc}
%\tabletypesize{\small}
\tablenum{3}
\tablewidth{0pt}
\tablecaption{SUMMARY OF AIT PHOTOMETRIC OBSERVATIONS XO-2N AND XO-2S}
\tablehead{
\colhead{Star} & \colhead{Observing} & \colhead{} & \colhead{Date Range} & 
\colhead{$\sigma$} & \colhead{Period} & \colhead{Full Amplitude} \\
\colhead{} & \colhead{Season} & \colhead{$N_{obs}$} & 
\colhead{(HJD $-$ 2,400,000)} & \colhead{(mag)} & \colhead{(days)} &
\colhead{(mag)} %\\
%\colhead{(1)} & \colhead{(2)} & \colhead{(3)} & \colhead{(4)} & 
%\colhead{(5)} & \colhead{(6)} & \colhead{(7)}
}
\startdata
 XO-2N  & 2013--2014 & 171 & 56572--56808 & 0.0035 & $29.89\pm0.16$ & $0.0049\pm0.0007$ \\
 XO-2N  & 2014--2015 &  97 & 56945--57154 & 0.0027 & $27.34\pm0.21$ & $0.0035\pm0.0007$ \\
 XO-2S  & 2013--2014 & 174 & 56572--56813 & 0.0025 &     \nodata    &       \nodata     \\
 XO-2S  & 2014--2015 & 105 & 56945--57154 & 0.0023 &     \nodata    &       \nodata     \\
\enddata
\end{deluxetable*}

%\clearpage
\begin{figure}
%\figurenum{5}
\epsscale{0.8}
\plotone{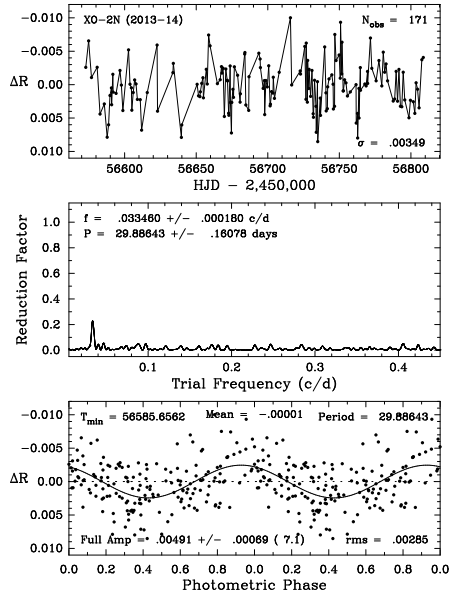}
\caption{{{$Top$: 171 nightly Cousins $R$-band photometric observations of 
XO-2N from the 2013--2014 observing season, acquired with the C14 Automated 
Imaging Telescope at Fairborn Observatory. $Middle$:  Frequency spectrum of 
the nightly observations finds a best period of $29.89\pm0.16$ days.  
$Bottom$: The observations phased with the best period give a full 
amplitude of $0.0049\pm0.0007$ mag, a 7$\sigma$ result.}}}
\label{fig:monitor}
\end{figure}

%\clearpage
\begin{figure}
%\figurenum{6}
\epsscale{0.8}
\plotone{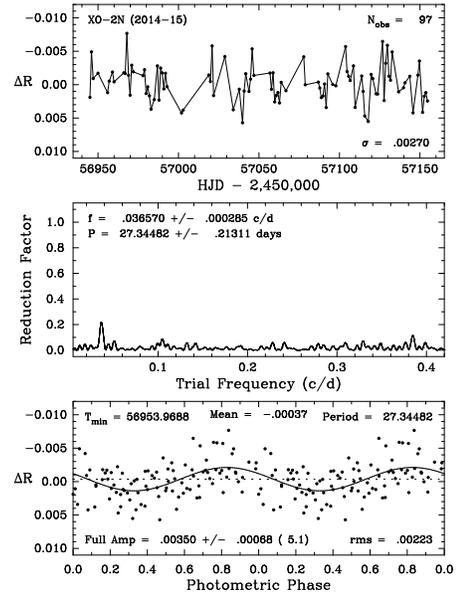}
\caption{{{$Top$: 97 nightly Cousins $R$-band photometric observations of 
XO-2N from the 2014--2015 observing season. $Middle$:  Frequency spectrum of 
the nightly observations finds a best period of $27.34\pm0.21$ days.  
$Bottom$: The observations phased with the best period give a full 
amplitude of $0.0035\pm0.0007$ mag, a 5$\sigma$ result.}}}
\label{fig:monitor14}
\end{figure}

%\clearpage
\begin{figure}
%\figurenum{7}
\epsscale{0.8}
\plotone{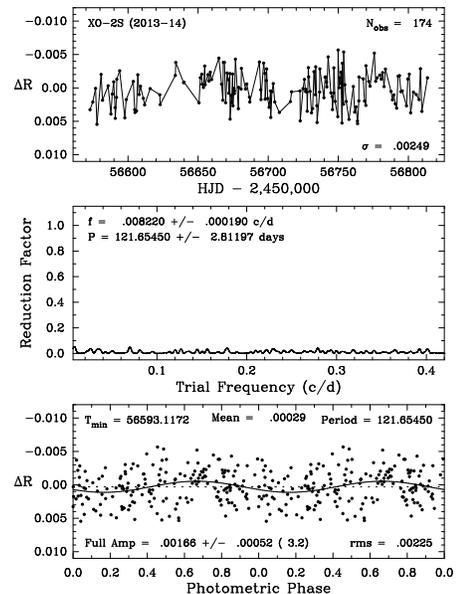}
\caption{{{$Top$: The nightly Cousins $R$-band photometric observations of 
XO-2S from the 2013--2014 observing season. $Middle$: Frequency spectrum of 
these observations finds no significant periodicity between 2 and 200 days. 
$Bottom$: Phase plot shows only a spurious period for the southern component 
in the 2013--2014 observing season.}}}
\end{figure}

%\clearpage
\begin{figure}
%\figurenum{8}
\epsscale{0.8}
\plotone{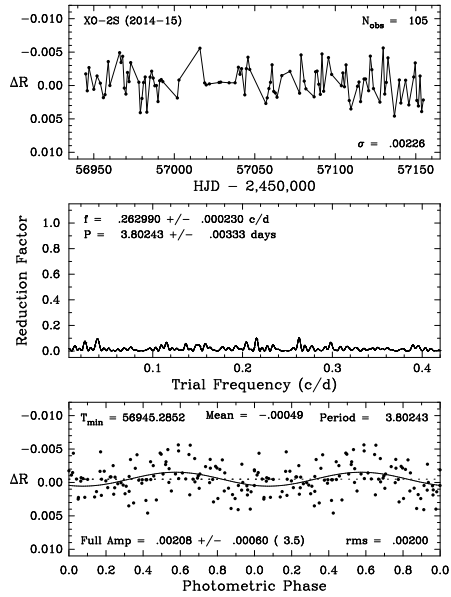}
\caption{{{$Top$: The nightly Cousins $R$-band photometric observations of 
XO-2S from the 2014--2015 observing season. $Middle$:  Frequency spectrum of 
the nightly observations finds no significant periodicity between 2 and 
200 days. $Bottom$:  Phase plot shows only a spurious period for the southern 
component in 2014--2015.}}}
\end{figure}

\begin{figure}%[htbp]
\begin{center}
%\vskip 0.2
%\epsscale{1}
%\plotone{XO-2-var.ps} 
%\includegraphics[width=0.9\columnwidth]{XO-2-var.ps}
\includegraphics[width=0.9\columnwidth]{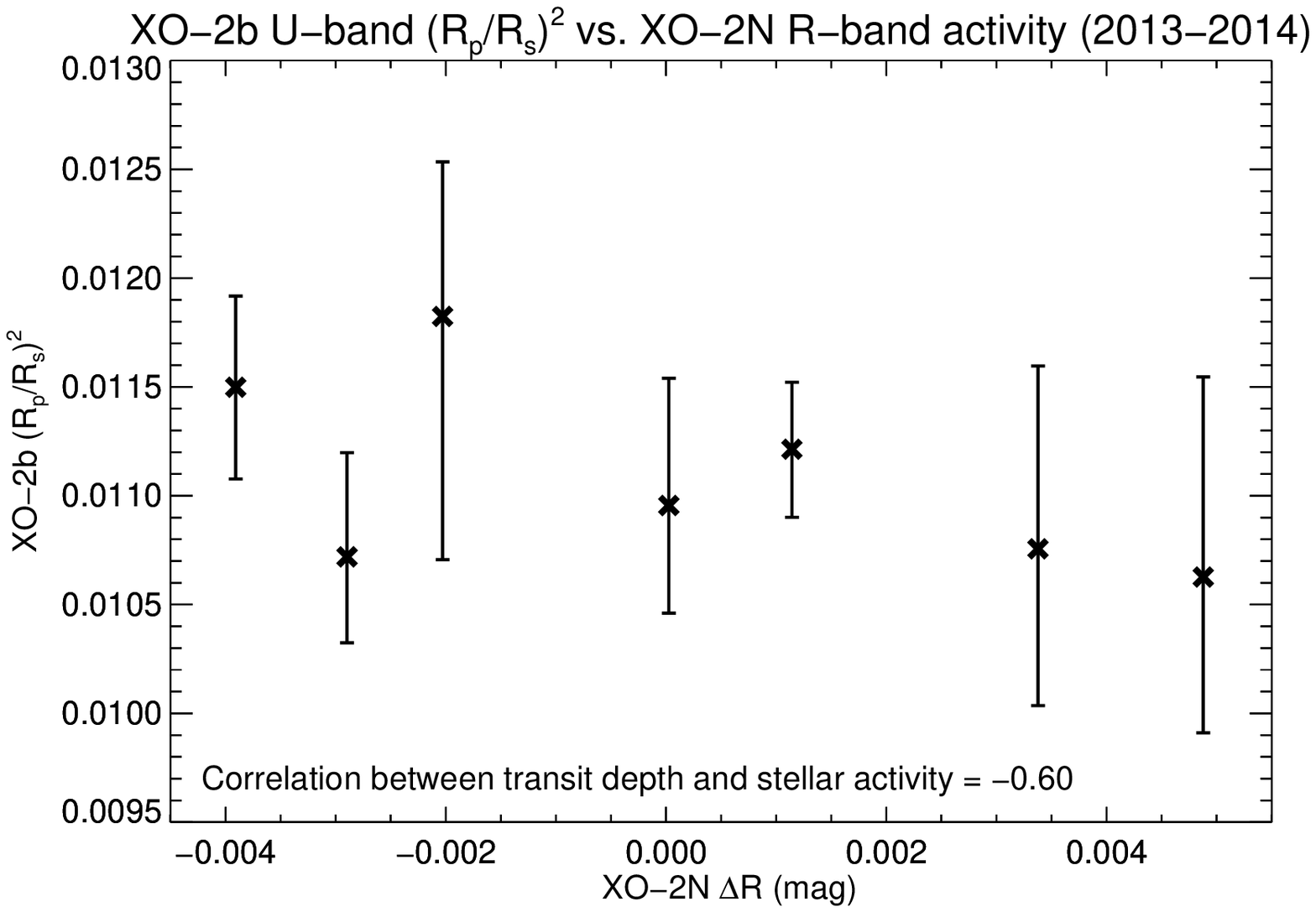}
\includegraphics[width=0.9\columnwidth]{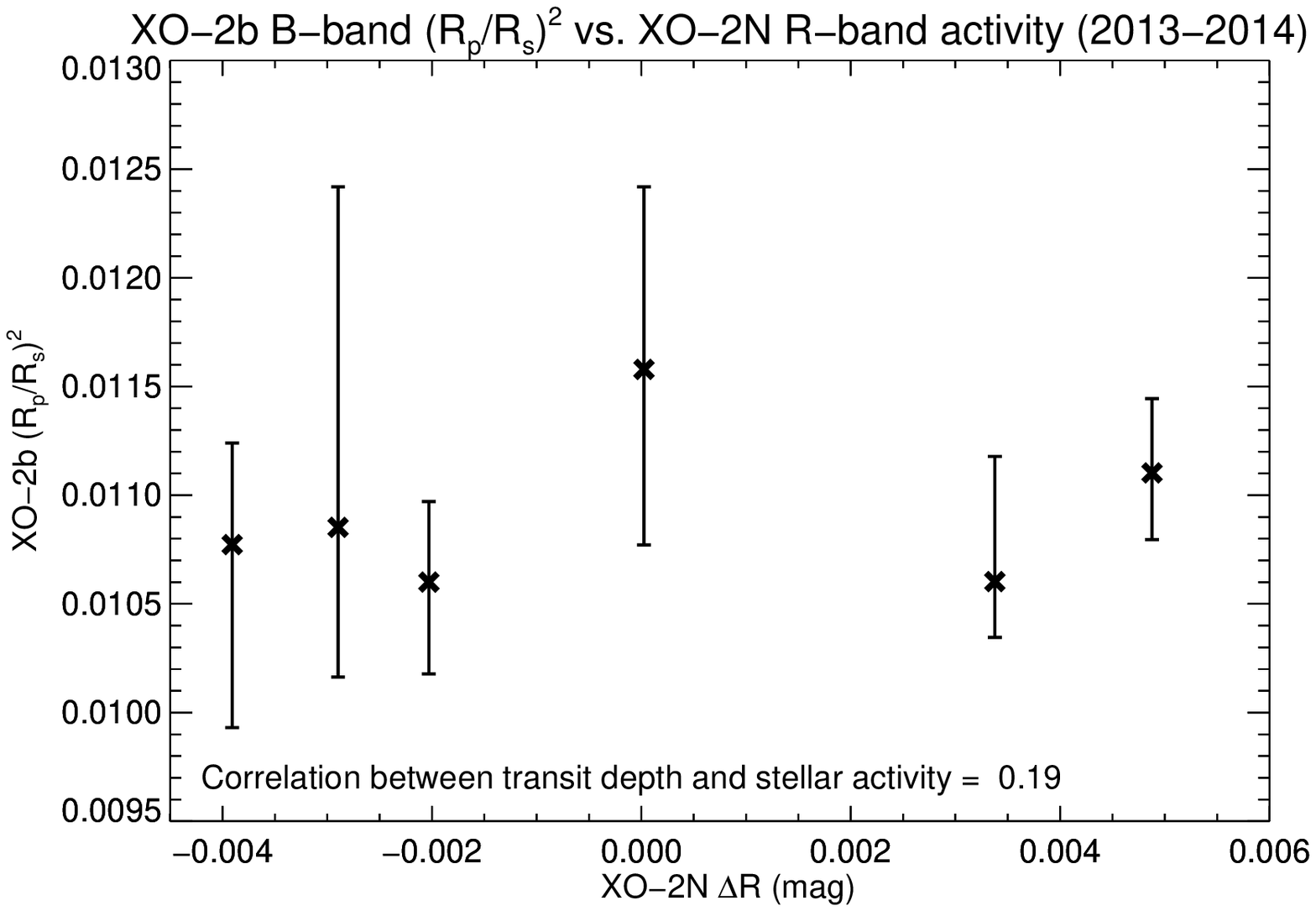}
\caption{{{XO-2b's Bessel U (top) and Harris B (bottom) photometric light curve depths (R$_{p}$/R$_{s}$)$^2$ recorded by the Kuiper 61'' telescope from 2012--2015 vs. XO-2N's $R$-band variability extrapolated from the AIT monitoring over the 2013--2014 season (Fig.~\ref{fig:monitor}). These plots suggest a correlation between the measured transit depth and host star variability in the U-band ($-0.60$) but not in the B-band ($0.19$). Since XO-2N's variability is likely affecting the derived transit depth in both passbands, we cannot bin the multiple nights of data together to achieve higher precision.} 
}}
\label{fig:variable}
\end{center}
\end{figure}

\begin{figure}%[htbp]
\begin{center}
%\vskip 0.2
%\epsscale{1}
%\plotone{XO-2-var.ps} 
%\includegraphics[width=0.9\columnwidth]{XO-2-var.ps}
\includegraphics[width=0.9\columnwidth]{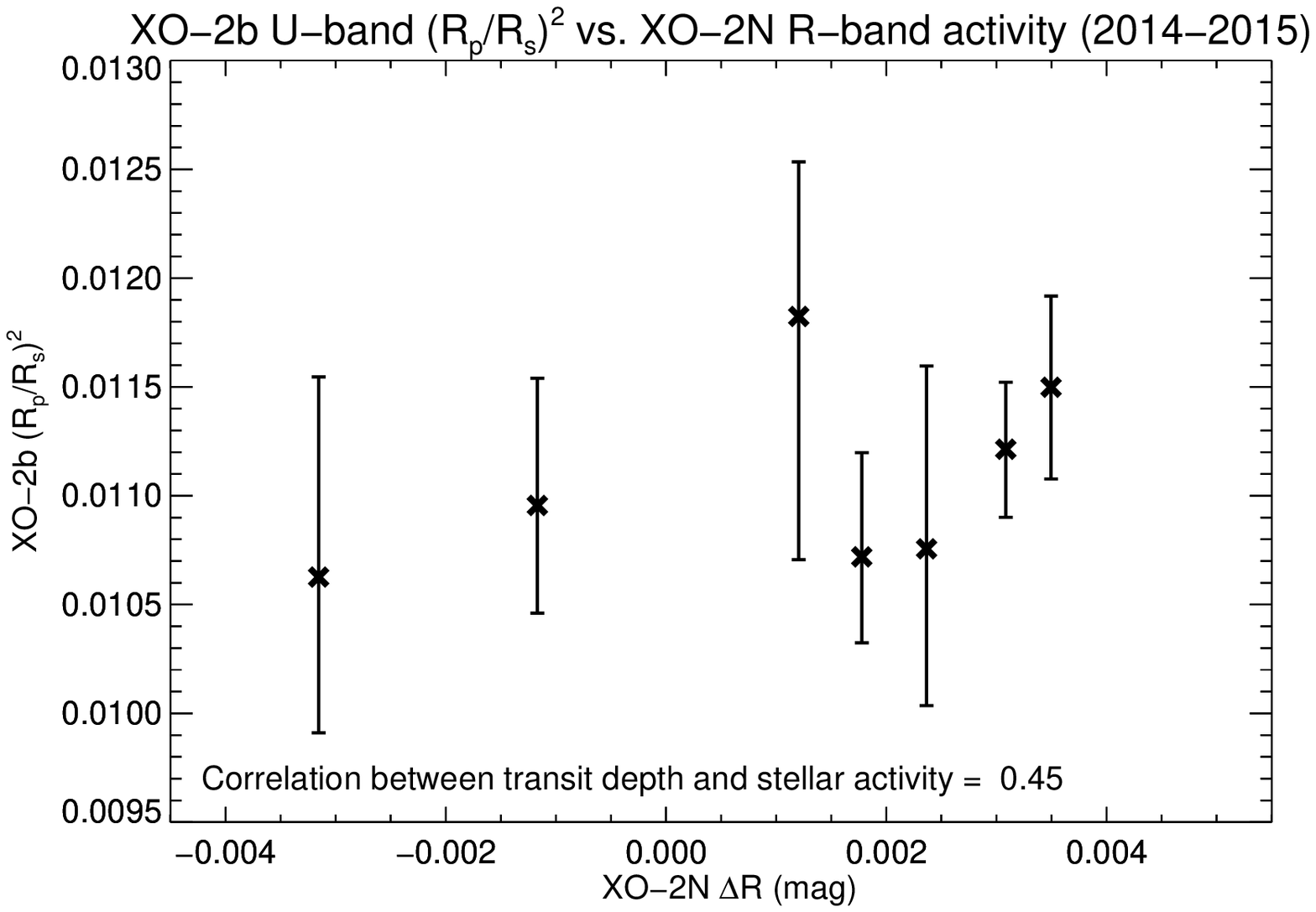}
\includegraphics[width=0.9\columnwidth]{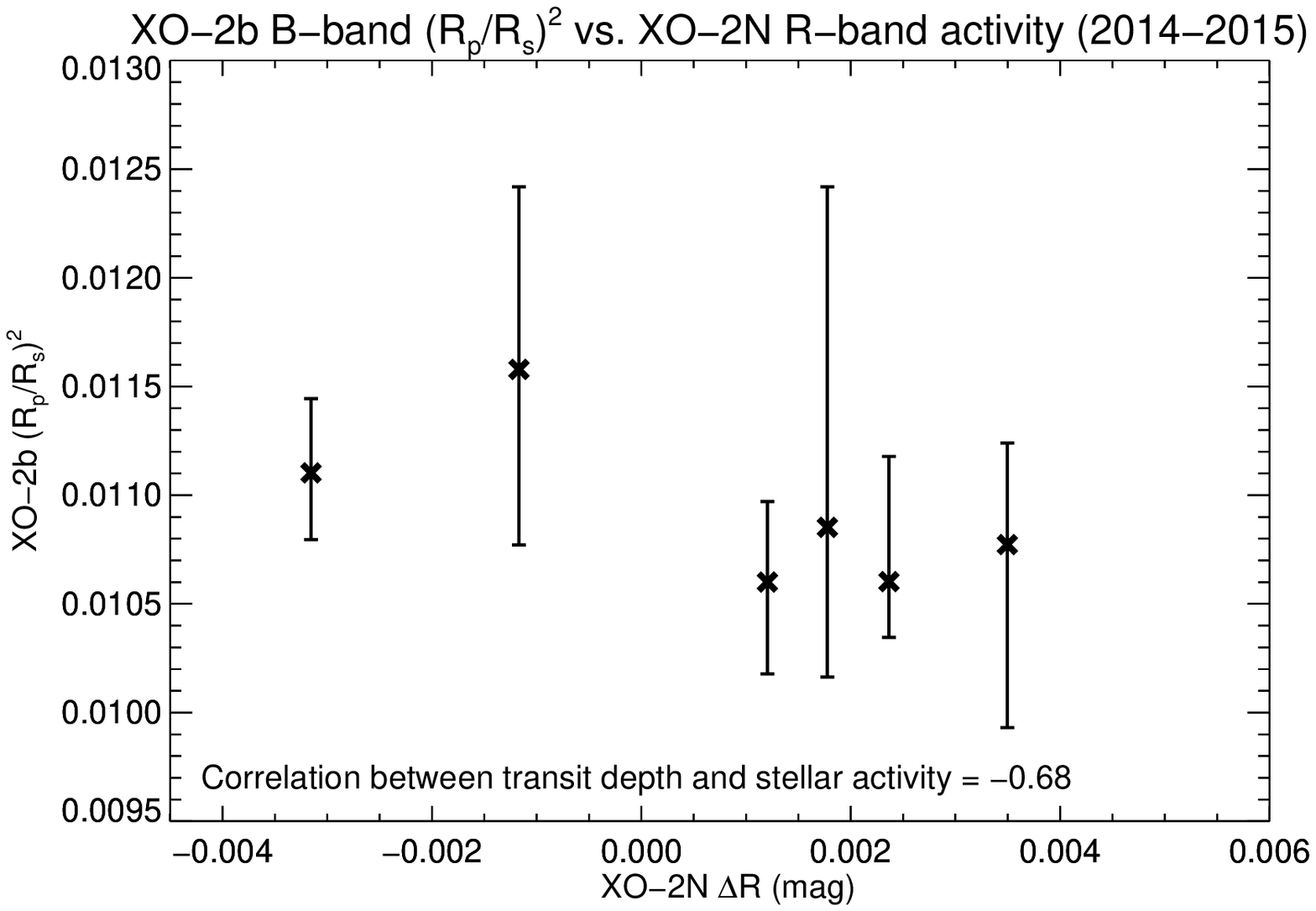}
\caption{{{XO-2b's Bessel U (top) and Harris B (bottom) photometric light curve depths (R$_{p}$/R$_{s}$)$^2$ recorded by the Kuiper 61'' telescope from 2012--2015 vs. XO-2N's $R$-band variability extrapolated from the AIT monitoring over the 2014--2015 season (Fig.~\ref{fig:monitor14}). These plots suggest a correlation between the measured transit depth and host star variability in both the U-band ($0.45$) and the B-band ($-0.68$).}}}

\label{fig:variable14}
\end{center}
\end{figure}

\bigskip
\section{Interpretation} 
\begin{figure}%[htbp]
\begin{center}
%\epsscale{1.18}
%\plotone{XO2trans-allUBstudiesb.eps} 
%\includegraphics[width=0.9\columnwidth]{XO2trans-allUBstudiesb.eps}
\includegraphics[width=0.9\columnwidth]{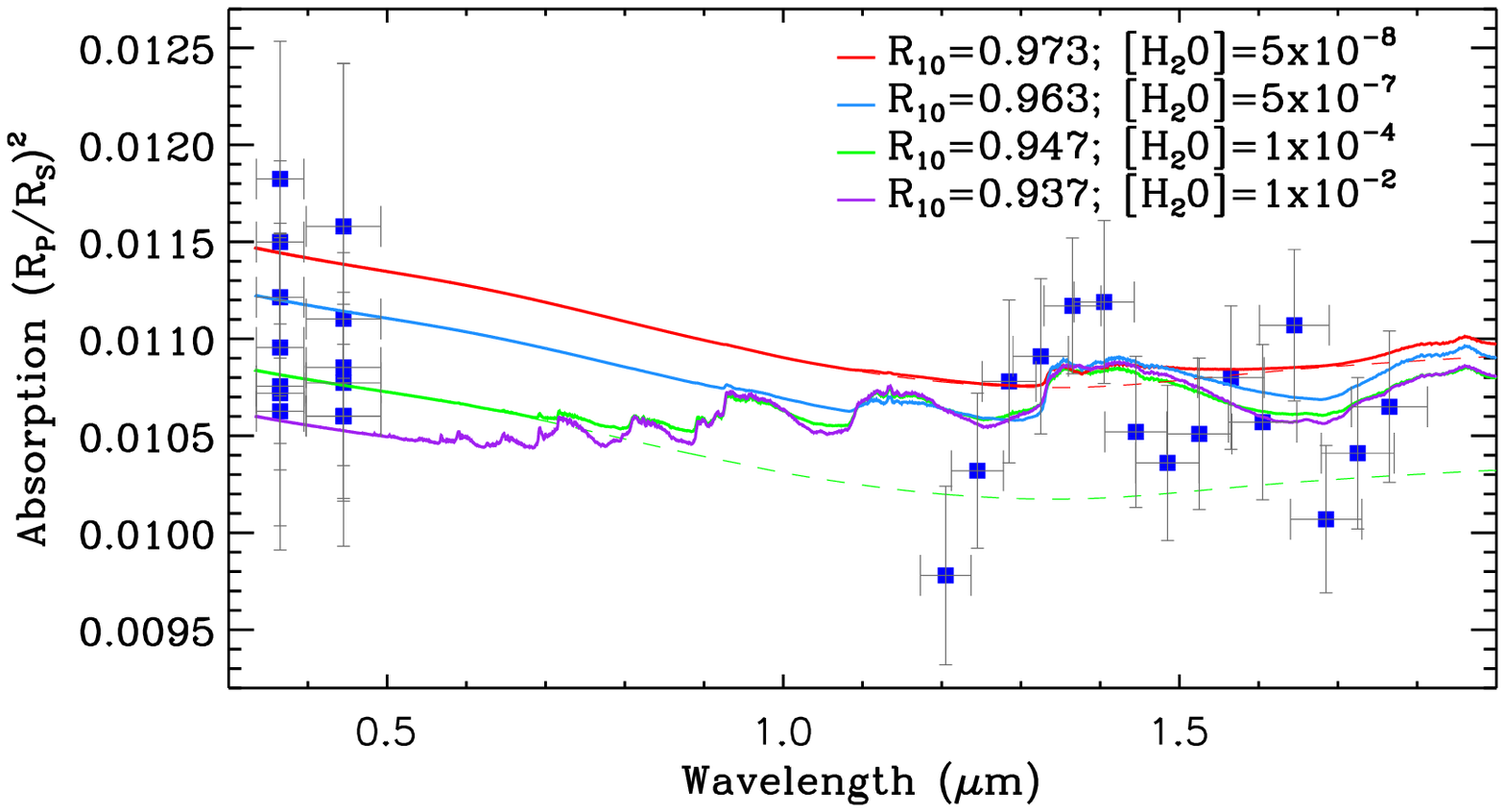}
\caption{$Hubble$/NICMOS measurements \citep{Crouzet12} of XO-2b's primary transit 
spectrum (1.2--1.8~$\micron$) and visible photometry from the 61'' Kuiper Telescope (0.3--0.5~$\micron$) are compared to calculated 
model spectra. Due to the variance in the transit depths measured both with $Hubble$ and the 61'', there is not a single radiative transfer model that can interpret all of the data simultaneously, preventing us from determining XO-2b's 10~bar radius (R$_{10}$) and placing strong constraints on its water abundance [H$_{2}$O] \citep[e.g., see][]{Griffith14a}.
}
\label{fig:Rdegen2} 
\end{center}
\end{figure}

To assess the effects of disparate values of the measured 
radius, we consider several radiative transfer models of XO-2b's data. 
We include Rayleigh scattering, absorption due to water and pressure-induced hydrogen absorption. The atmosphere is assumed to be cloudless. Since XO-2b has an equilibrium temperature between HD~189733b and HD~209458b, we adopt a thermal profile intermediate between those of HD~189733b and HD~209458b \citep{Moses11}. The line parameters for H$_2$O are calculated from the parameters of 
\citet{Rothman09,Tashkun03,Barber06}, while 
hot pressure-induced H$_2$ absorption derives from \citet{Borysow02,Zheng95}. As shown in Figure~\ref{fig:Rdegen2}, uncertainties in the XO-2b's U and B-band light curve depths lead to strong uncertainties in the
pressure-dependent radius, which in turn leads to poor constraints on the composition, even
in the event of highly precise near-IR measurements, currently possible with $Hubble$/WFC3 \citep{Griffith14a}. Our observations of XO-2b indicate that its host star variability prevents us from combining data from multiple nights in order to reduce our uncertainties. These limitations indicate the great need for measurements on different platforms, monitoring programs with one instrument, and simultaneous observations at different wavelengths, which  
sample the same effective radius (and can then be interpreted together).

%The results of our analysis are shown in Figure~\ref{fig:Rdegen2}.  As found by 
%Griffith (2014), the derived mixing ratios of water are strongly dependent 
%on the assumed radius such that variations at the level of 1\% in the radius
%causes uncertainties of an order of magnitude. 
%This study is indicative of an approach rather than 
%a precise measurement of XO-2b's atmosphere, because there are 
%discrepant data points, large errors, few measurements and variable 
%stellar activity. This variability prevents us from combining data together from multiple nights in order to reduce our uncertainties. These limitations indicate the great need for 
%measurements on different platforms, monitoring programs on one instrument,  and simultaneous observations at different wavelengths, which  
%sample the same effective radius (and can then be interpreted together).

%since the effects from other predicted abundant molecules are expected to be minimal. 
%In addition, we assume  and a cloudless atmosphere.

\section{Conclusions}
%Here we present a four year study to determine the transiting hot Jupiter XO-2b's planet-to-star radius ratio in the photometric U and B-bands in order to measure the Rayleigh scattering properties of its atmosphere.

{{Our observations indicate that XO-2b's host star, XO-2N, is variable with a peak-to-peak amplitude of $0.0049 \pm 0.0007$~mag in R and a $29.89 \pm 0.16$~day period in the 2013--2014 season and a peak-to-peak amplitude of $0.0035 \pm 0.0007$~R-mag and a period of $27.34 \pm 0.21$~days for the 2014--2015 observing season. This host star variability potentially influences XO-2b's transit depths.}} Due to the stellar variability, our data cannot be binned to reduce uncertainties. Our study not only demonstrates that ground-based monitoring of a host star for variability is crucial for transit observations, but also that visible and near-IR observations must be taken at the same time to insure that the same stellar conditions are probed.

% While we lack the precision here to conclusively determine how much an effect XO-2N's variability is having on the transit depths, we predict that larger telescopes ($\gtrsim$3~m) have the necessary sensitivity. 

\acknowledgments{The work by R. T. Zellem, C. A. Griffith, K. A. Pearson, J. D. Turner, and J. K. Teske was supported by the NASA Atmospheres Program.

{{GWH and MHW thank Lou Boyd for his many years of APT support at
Fairborn Observatory and acknowledge support from NASA, NSF, Tennessee
State University, and the State of Tennessee through its Centers of 
Excellence program.}}

{{RTZ thanks Mario Damasso, Nikole K. Lewis, Mahmoudreza Oshagh, and Mark R. Swain for their helpful discussions.}}

We thank the anonymous referee for their helpful comments and suggestions.
}

\bibliographystyle{apj}       %need ApJ bst file
\bibliography{cgexo}

%%  Run bibtex once, then latex Eq-Clouds8 twice

\end{document}